\documentclass[aps,pra,twocolumn,superscriptaddress,showpacs,amsmath,amstex,amssymb,citeautoscript]{revtex4-1}

\usepackage[T1]{fontenc}
\usepackage[utf8]{inputenc}
\usepackage{lipsum}
\usepackage{amsmath}
\usepackage{amssymb}
\usepackage{bbm}
\usepackage{braket}
\usepackage{xcolor}
\usepackage{pifont}
\usepackage[mathscr]{euscript}
\usepackage[shortlabels]{enumitem}
\usepackage[justification=justified]{subcaption}
\usepackage{graphicx}
\usepackage{lipsum}
\allowdisplaybreaks
\usepackage{float}
\usepackage{graphicx}
\usepackage{dsfont}
\usepackage{comment}
\usepackage[colorlinks=true]{hyperref}  
\hypersetup{
    bookmarks=true,         
    unicode=false,          
    pdftoolbar=true,        
    pdfmenubar=true,        
    pdffitwindow=false,     
    pdfstartview={FitH},    
    pdftitle={Correlations between superconducting and resistive anisotropies},    
    pdfauthor={Banerjee, Scammell, Scheurer},     
    pdfsubject={},   
    pdfcreator={},   
    pdfproducer={}, 
    pdfkeywords={} {} {}, 
    pdfnewwindow=true,      
    colorlinks=true,       
    linkcolor=blue, 
    citecolor=blue,        
    filecolor=magenta,      
    urlcolor=blue           
}

\newcommand{\equref}[1]{Eq.~(\ref{#1})}

\newcommand{\secref}[1]{Sec.~\ref{#1}}
\newcommand{\figref}[1]{Fig.~\ref{#1}}
\newcommand{\refcite}[1]{Ref.~\onlinecite{#1}}

\newcommand{\tableref}[1]{Table~\ref{#1}}
\newcommand{\appref}[1]{Appendix~\ref{#1}}
\newcommand{\mc}{\mathcal}

\newcommand{\pdagger}{{\phantom{\dagger}}}

\newcommand{\sign}{\,\text{sign}}

\renewcommand{\vec}[1]{\boldsymbol{#1}}

\definecolor{wrongultramarine}{rgb}{1,0.5,0}

\begin{document}

\title{Correlations between superconducting and resistive anisotropies}

\author{Sayan Banerjee}
\affiliation{Institute for Theoretical Physics III, University of Stuttgart, 70550 Stuttgart, Germany}

\author{Harley D.~Scammell}
\affiliation{School of Mathematical and Physical Sciences, University of Technology Sydney, Ultimo, NSW 2007, Australia}

\author{Mathias S.~Scheurer}
\affiliation{Institute for Theoretical Physics III, University of Stuttgart, 70550 Stuttgart, Germany}

\begin{abstract}
There are multiple possible origins of transport anisotropies in metals and superconductors. For instance, rotational symmetry can be spontaneously broken in the normal state as a result of electronic nematic order inducing anisotropies in an otherwise $s$-wave superconducting phase. Another possibility is that the dominant source of rotational symmetry breaking is the superconductor itself and its vestiges that may survive in the normal state.
We here theoretically analyze the correlations of transport anisotropies in the normal and the corresponding superconducting phase for different scenarios of broken symmetry, either coming solely from the normal state, solely from the superconductor and its vestiges in the metallic regimes, or from both simultaneously. We further include both zero-momentum and finite-momentum pairing; we develop a theory of vestigial order for the latter, characterized by broken rotational and translational symmetry. Our findings reveal that the relative transport anisotropies in the normal and superconducting phases sensitively depend on the scenario, including the form of vestigial order and, in some cases, the parity of the superconducting order parameter. As such, measuring the directional dependence of the critical current and resistivity can provide strong constraints on the origin of rotational symmetry breaking. We demonstrate our findings in minimal models relevant to twisted multilayer graphene, rhombohedral graphene, and twisted transition metal dichalcogenides. 
\end{abstract}

\maketitle

\section{Introduction}
Symmetries are a key guiding principle for our understanding of both metallic and superconducting states~\cite{sigrist_unconventional}. In weakly interacting metals, transport properties primarily reflect the symmetries of the crystalline lattice. In contrast, for strongly interacting systems, electronic correlations can drive additional symmetry breaking transitions that do not stem from the lattice itself. In the case of electronic nematic order \cite{nem_frad,nematic_rev}, only rotational symmetry is broken while translational symmetry is preserved. However, rotational symmetries can also be spontaneously broken by the formation of charge-density waves (CDWs) where the charge is modulated only along one direction. 
When a superconductor (SC) emerges from such a symmetry-reduced electron liquid, the superconducting condensate will inherit the same (broken) symmetries, resulting in anisotropic superconducting properties. However, the symmetry reduction in the normal state can also stabilize pairing states that would otherwise not be favored energetically, since, e.g., a nematic superconducting order parameter can couple to the first power of the nematic order parameter of the normal state. 
Beyond spatial symmetries, also the breaking of time reversal symmetry (TRS) in the normal state has profound consequences for pairing; the degeneracy of states with momenta $\vec{k}$ and $-\vec{k}$ is lifted which can lead to non-zero momentum of Cooper pairs \cite{FF,LO} possibly lowering the system's symmetries further \cite{stripedBerg,berg2009charge, PDW}. 

Conversely, superconductors themselves can also spontaneously break symmetries \cite{sigrist_unconventional}, and, upon melting of the superconducting order, these broken symmetries can imprint on the normal-state responses through ``vestigial orders'' \cite{kivelson1998electronic,fernandes2019intertwined}---a very actively studied field at the moment \cite{grinenkoStateSpontaneouslyBroken2021,geDiscoveryCharge4eCharge6e2022,fernandesChargeSuperconductivityMulticomponent2021,PhysRevB.85.024534,jianChargeSuperconductivityNematic2021,zengPhasefluctuationInducedTimeReversal2021,songPhaseCoherencePairs2022,PhysRevB.107.064501,chungBerezinskiiKosterlitzThoulessTransitionTransport2022,jiangCharge4eSuperconductors2017,LI20242328,gnezdilovSolvableModelCharge2022,garaudEffectiveModelMagnetic2022,Sixfermions,hecker_vestigial_2018,PhysRevB.108.054517,zhouChernFermiPocket2022,poduval_vestigial_2024,2024npjQM...9...66W,2024PhRvL.133m6002L,PhysRevB.110.054519,ScammellVestigialKagome,2025arXiv251002474V,PhysRevB.109.134514,PhysRevB.107.224503,PhysRevB.109.144504,Volovik4e,2025arXiv251005209H}.  
All of this suggests a deep and intricate relationship among symmetries in the normal states and superconducting phases. 

One of the central goals in the field of superconductivity is to identify the form and symmetries of the superconducting order parameter in a given material---in most cases, however, this is very challenging. For instance, ever since the discovery of superconductivity in two-dimensional (2D) twisted-graphene superlattices \cite{cao2018unconventional,yankowitz2019tuning,park2021tunable,park2022robust,oh2021evidence}, which can exhibit a plethora of conventional and unconventional states, the mechanism and type of pairing have been subjects of intense debate, see, e.g., \cite{lake, waschitz2025momentumresolvedspectroscopysuperconductivityquantum,PhysRevB.98.220504,PhysRevB.98.241407,PhysRevResearch.2.033062,PhysRevX.8.041041,BODPairingNatComm,PhysRevB.99.195114,PhysRevLett.121.257001,huang2019antiferromagnetically,PhysRevB.99.144507,cea, PhysRevLett.127.247703, PhysRevB.103.024506,PhysRevB.111.L020502,PhysRevLett.133.146001,dassarma,shavit2025nematicenhancementsuperconductivitymultilayer} and many more. Recently, other (untwisted) rhombohedral graphene stacks \cite{SCTrilayer,rtg_chiral_sc_experiment} have also surged in interest, and multiple exotic candidate pairing states have been discussed \cite{rhomohedral1,yoon2025quartermetalsuperconductivity,maymann2025pairingmechanismdictatestopology,wy3f-hgr9,geier2024chiraltopologicalsuperconductivityisospin,2024arXiv241108969S,yang2024topologicalincommensuratefuldeferrelllarkinovchinnikovsuperconductor,qin2025chiralfinitemomentumsuperconductivitytetralayer,jahin2025enhancedkohnluttingertopologicalsuperconductivity,2025arXiv250305697M,christos2025finitemomentumpairingsuperlatticesuperconductivity,sedov2025probingsuperconductivitytunnelingspectroscopy,gil2025chargepairdensitywaves,zfmh-rjzc,k8sb-rqxf}.  

One particular avenue to probe the pairing symmetry is based on transport anisotropies in both the normal (resistivity) and superconducting states (critical current). 
For instance, in magic-angle twisted bilayer graphene, a twofold anisotropy in the
resistivity and critical current was observed and attributed to be a signature of nematic pairing \cite{cao2021nematicity}. 
However, rotational symmetry breaking in the form of external strain \cite{strain1,strain2,strain3} or nematicity has also been shown to exist in the normal state \cite{kerelsky2019maximized,choi2019electronic,stepanov2020untying,jiang2019charge,cao2021nematicity}, which might or might not be related to superconductivity. It is quite hard to disassociate whether the SC breaks rotational symmetry explicitly or inherits it from the normal state \cite{PhysRevResearch.2.033062,nem_sc1,nem_sc2,nem_sc3,nem_sc4,nem_sc5,shavit2025nematicenhancementsuperconductivitymultilayer}. What is more, the point group often allows for more than one possible nematic superconducting state, and it would be interesting to determine whether, and under which conditions, transport anisotropies can be used to distinguish even-parity and odd-parity nematic superconductors. 

Another interesting experimental finding in twisted graphene superlattices
was reported in \cite{zhang2025angularinterplaynematicitysuperconductivity}, where an atypical behavior was observed in the directionalities of the superconducting and normal states. Specifically, a transport anisotropy was observed where the direction of the maximum critical current in the superconducting state coincided with that of the maximum resistivity in the normal state. A Ginzburg-Landau analysis was performed, which revealed that if one takes an anisotropic normal state and a uniform gap, one obtains an opposite relationship between the two, which is in stark contrast to the experimental observation. While this places constraints on the order parameter, one can also imagine a situation where fluctuation-driven vestigial orders play a role, going beyond strain-dominated effects. Additional important open aspects are the role of trigonal warping and that additional anisotropies in the normal state, which are unrelated to superconductivity, will also affect the superconducting energetics itself and select specific order-parameter configurations.

Motivated by the rich physics and sensing capabilities of correlations between normal and superconducting angular-dependent transport, we here study various possible cases and sources---in particular, those relevant to graphene and transition metal dichalcogenide (TMD) heterostructures---of transport anisotropies in both normal and superconducting phases. We demonstrate the interplay of various primary and vestigial orders and their transport signatures, both in mean-field theory and by taking fluctuations into account. 

The presence of $U(1)_{v}$ valley symmetry in these systems allows us to discuss the sources of rotational symmetry breaking for two kinds of pairing states separately: inter- and intravalley superconductivity. We perform an in-depth analysis for both cases. 
For the intervalley case (see \secref{Intervalley}), we first outline the general formalism to evaluate the angular dependencies of critical supercurrents and resistivities. We then show the dependencies for $\vec{k}$-independent superconducting order parameters and for nematic pairing, considering also the case where the superconducting order parameter or fluctuations thereof are the only sources of rotational symmetry breaking. By comparing the interplay of the anisotropies, we show that there are features which can be used to extract the parity of the order parameter and gain crucial insights into the nature of superconductivity. In \secref{Intravalley}, we shall consider the intravalley pairing case. We develop a theory of vestigial orders emerging out of finite-momentum superconductivity in a single valley, which is particularly relevant to multilayer rhombohedral graphene. We will see how this can lead to striped and nematic phases \cite{murshed2025chargedensitywavesstripesquarter,morissette2025stripedsuperconductorrhombohedralhexalayer,nguyen2025hierarchysuperconductivitytopologicalcharge}, emerging out of the primary order parameter provided by superconductivity.
Our analysis will show how different forms of orders (all breaking the same symmetries) can manifest differently in transport. While we broadly base our analysis on van der Waals heterostructures, we emphasize that our formalism and results are also directly applicable to other systems such as kagome materials; in particular, an anti-correlation between normal state and superconducting anisotropic transport behavior has been reported in the kagome superconductor CsV$_3$Sb$_5$ \cite{kagomeanisotropy}.


\section{Intervalley pairing}\label{Intervalley}
\subsection{General Formalism}\label{sec:generalforma}
Before presenting explicit calculations, we start by discussing the general methodology to extract normal state and resistive properties for intervalley-paired SCs. We shall first focus on the case where the normal state has TRS and the band dispersions in the two valleys, $\nu=\pm$, are just mirror images of one another, related by $\vec{k} \rightarrow -\vec{k}$.
Although the generalization to include spin is straightforward, we will, for the simplicity of the presentation, assume that either the Fermi surfaces are spin polarized or spin and valley are locked. This applies to twisted TMDs, such as twisted WSe$_2$ \cite{WSe2SC1,WSe2SC2}, as a result of the strong spin-orbit coupling and is further natural for the spin polarized half-metal phases in rhombohedral graphene \cite{rtg_chiral_sc_experiment}, twisted MoTe$_2$ \cite{MoTe2_unconv_exp}, or for pairing around filling fraction $2$ in twisted multi-layer graphene \cite{PauliLimit,morissetteElectronSpinResonance2022}. 
As a result of TRS, pairing between opposite valleys and momenta $\vec{k}$ and $-\vec{k}$ is favored, and the superconducting order parameter has to transform under an irreducible representation (IR) of the normal-state point group. While interband pairing can lead to interesting superconducting properties \cite{BODPairingNatComm,PutzerBOD}, we will here focus on IRs with band-diagonal order parameters. Then the relevant active low-energy electrons can be described by the creation operators $c^\dagger_{\vec{k},\nu}$ and the pairing Hamiltonian is given as 
\begin{align}\begin{split}
    \mathcal{H}_c = g_{c}\sum_{\vec{k}}\sum_{j=1}^{d_{\Gamma}} \left[ \eta_{j}(\vec{q})f_{j}(\vec{k}) c^\dagger_{\vec{k}+\vec{q},+}c^\dagger_{-\vec{k}+\vec{q},-}  + \text{H.c.} \right].\label{InterV}
\end{split}\end{align} 
Here, $j$ runs over $d_{\Gamma}$ components of the IR $\Gamma$ with form factors $f_j(\vec{k})$. For the point groups relevant to us here, only $d_{\Gamma}=1,2$ are possible and we will discuss examples with both $d_{\Gamma}=1$ and $d_{\Gamma}=2$ in this work. 

As explained above, our goal is to compare transport anisotropies in the superconducting and normal state. As such, we will consider normal-state Hamiltonians, $\mathcal{H}_0=\sum_{\vec{k},\nu=\pm} c^\dagger_{\vec{k},\nu} E_{\vec{k},\nu}^{\pdagger} c^\pdagger_{\vec{k},\nu}$, where $C_{3z}$ rotational symmetry is broken. The associated dispersion can take either of the following forms:
\begin{subequations}\begin{align}
    E_{\vec{k},\nu} &=\xi_{\vec{k},\nu} + \vec{\Phi} \cdot \vec{g}(\vec{k}) \quad \text{or} \label{NematicStrain} \\
    E_{\vec{k},\nu} &=\xi_{\vec{k},\nu} + \delta\epsilon_{\vec{k},\nu}. \label{VestigialBrokenSyms} 
\end{align}\end{subequations}
Here $\xi_{\vec{k},\nu} = \epsilon_{\nu\cdot\vec{k}}$, such that $\xi_{\vec{k},\nu} = \xi_{-\vec{k},-\nu}$, is the single-particle dispersion for valley $\nu$, which obeys $\epsilon_{C_{3z}\vec{k}}=\epsilon_{\vec{k}}$. This first form (\ref{NematicStrain}) describes symmetry breaking as a consequence of moiré nematic order or effectively the presence of strain \cite{kerelsky2019maximized,choi2019electronic,stepanov2020untying,jiang2019charge,cao2021nematicity,2025arXiv250510506M}, which we will parameterized by $\vec{\Phi} = \Phi (\cos{\theta},\sin{\theta})$; meanwhile, \equref{VestigialBrokenSyms} models the consequence of a vestigial order $\delta\epsilon_{\nu,\vec{k}}$ that modifies the bare dispersion. These are the two principal sources of breaking the symmetry that we will discuss in this work. To preserve TRS (and, if present, $C_{2z}$ symmetry), we choose $\vec{g}(\vec{k})$ to be even in momentum $\vec{k}$ and to transform as a vector under $C_{3z}$ (specifically IR $E$ of the point group $D_{3}$ or $E_2$ of $D_6$). In the limit of small $\vec{k}$, its leading behavior is given by $\vec{g}(\vec{k}) \sim (k_y^2 -k_x^2,2k_xk_y)$ and we take the Brillouin-zone-periodic extension with the fewest number of nodal lines. 

In order to probe such $C_{3z}$ symmetry breaking in experiments, one typically measures the resistivity/conductivity tensor, which is sensitive to such anisotropies. Within a Boltzmann transport approach, the conductivity tensor in the relaxation-time approximation is given as (see e.g. \cite{rhine_Nematic} where a similar notation is used)
\begin{subequations}\begin{equation}
\sigma^\pdagger_{\alpha\beta} = \frac{2e^2 \tau}{\mc{V}} \sum_{\nu=\pm} \left\langle(\vec{v}^\pdagger_{\vec{k},\nu})^\pdagger_\alpha (\vec{v}^\pdagger_{\vec{k},\nu})^\pdagger_\beta\right \rangle_\nu, \label{GeneralExpressionForSigma}
\end{equation}
where $(\alpha,\beta)\in \{x,y\}$ and all the nonuniversal properties are encoded in the generalized average
\begin{equation}
    \braket{\dots}_\nu \equiv \frac{4}{T N}\sum_{\vec{k}} \frac{\tau^\pdagger_{\vec{k},\nu}/\tau}{\cosh^2\hspace{-0.1em} \left(\frac{E_{\vec{k},\nu}-\mu}{2T}\right)} \dots \label{ExpectationValue}
\end{equation}\label{FullExpressionForSigma}\end{subequations}
and in the velocities $\vec{v}_{\vec{k},\nu}$\,$=$\,$ \vec{\nabla}_{\vec{k}} E_{\vec{k},\nu}$. Here, $N$ is the number of moir\'e unit cells, $\mc{V}$ is the area of the moir\'e unit cell, and $\tau_{\vec{k},\nu}^{-1}$ is the relaxation-time scattering rate for momentum $\vec{k}$ in valley $\nu$. The average scattering time is defined as $\tau$\,$=$\,$N^{-1} \sum_{\vec{k}} \tau_{\vec{k},\nu}$, which is the same in both valleys, due to TRS. Considering a current applied along
a direction that makes an angle $\Omega$ with the $x$ axis, and assuming a momentum-independent scattering time $\tau_{\vec{k},\nu} = \text{const.}$, one can now extract the angular dependence of longitudinal resistivity as
\begin{equation}
        \rho^\pdagger_L(\Omega)  = \frac{1}{\sigma_{0}} - \frac{1}{\sigma_{0}^2} \left( \lambda \cos 2\Omega +\delta \sin 2\Omega\right),
\end{equation}
where $\lambda = (\sigma_{xx}-\sigma_{yy})/2$, $\sigma_{0} = (\sigma_{xx}+\sigma_{yy})/2$ and $\delta= \sigma_{xy}$. 

Equipped with normal state transport, we now focus on the superconducting phase. Assuming that the SC arises from such a normal state, we take an attractive interaction ($g_c >0$) and perform a mean field decoupling to obtain $H_{s} = \mathcal{H}_{0} + \mathcal{H}_{c} + \sum_{\vec{q},j}|\eta_{j}(\vec{q})|^2/g_c$. Integrating out the electrons, and expanding up to quadratic order in the free energy, we get 
\begin{equation}
    \mathcal{F} \sim \sum_{\vec{q},j=1}^{d_{\Gamma}}a_{j}(\vec{q})|\eta_{j}(\vec{q})|^2 + \mathcal{O}(\eta^4).
    \label{eq:FreeEnergy}
\end{equation}
It holds $a_{j}(\vec{q}) = \frac{1}{g_c}-\Gamma^{j}(\vec{q})$ with
\begin{equation}
   \Gamma^{j}(\vec{q})  = \frac{1}{2N}\sum_{\vec{k}} \frac{\tanh{\frac{\mathcal{E}^{\Phi}_{\vec{k},\vec{q},+}}{2 T}} + \tanh{\frac{\mathcal{E}^{\Phi}_{\vec{k},\vec{q},-}}{2 T}}}{\mathcal{E}^{\Phi}_{\vec{k},\vec{q},+} + \mathcal{E}^{\Phi}_{\vec{k},\vec{q},-}}|f_{j}(\vec{k})|^2,\label{ParticleParticleBubble}
\end{equation}
where $N$ is the number of unit cells and $\mathcal{E}_{\vec{k},\vec{q},\nu}^{\Phi} = E^{\Phi}_{\vec{k}+\nu\vec{q},+}$.
Restricting the analysis to single-$\vec{q}$ states, $\eta(\vec{q}) \propto \delta_{\vec{q},\vec{q}_0}$, we can then obtain the current $\vec{J}(\vec{q}) =  2e\partial_{\vec{q}}\mathcal{F}$ . The critical current $J_c(\Omega)$ along a certain direction $\Omega$ is then given by the maximum of  $|\vec{J}(\vec{q})|$ with $\vec{J}(\vec{q})$ oriented along $\Omega$. 
To compare the relative anisotropies between the superconducting critical current and normal state resistivity, we define the quantity
\begin{equation}
    \zeta_{xy} := \frac{\text{sgn}[\bar{J}_{c}(\Omega=0)-\bar{J}_{c}(\Omega=\pi/2)]}{\text{sgn}[\bar{\rho}_{L}(\Omega=0)-\bar{\rho}_{L}(\Omega=\pi/2)]}. \label{DefinitionOfZeta}
\end{equation}
We here defined $\bar{J}_{c}(\Omega) := \text{max}_{n=0,1,2} J_{c}(\Omega + n 2\pi/3)$ and $\bar{\rho}_{L}(\Omega) := \text{max}_{n=0,1,2} \rho_{L}(\Omega + n 2\pi/3)$ to account for all three, $C_{3z}$-related, orientations of the nematicity.
By construction, $\zeta_{xy} \in \{-1,1\}$,  indicating whether the anisotropy of the critical current is aligned ($\zeta_{xy} = 1$) or anti-aligned ($\zeta_{xy} = -1$)  with that of the resistivity. While equivalent for many of our model calculations, we note that we use $\zeta_{xy}$ instead of stating whether the maxima of $J_{c}$ and $\rho_{L}$ are aligned or not; the reason is that in some cases, these quantities have their global maxima close to but not exactly at the axes $\Omega = 0$ and $\pi/2$ [where there are technically only local minima, pinned by reflection symmetries, see, e.g., \figref{fig:nem_sc2}(a)]. In those cases, $\zeta_{xy}$ is a more useful quantity to distinguish the behavior.  

\begin{figure}[tb]
    \centering
\includegraphics[width=\linewidth]{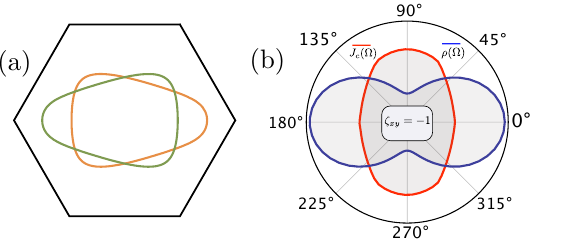}
    \caption{(a) Fermi surfaces (FSs) for the two-valley model. The presence of nematic order along $x$ ($\theta = 0 $) breaks the $C_{3z}$ symmetry in the normal state (b) shows the resistivity (in blue) and critical current (in red) for $\Phi = 0.2$. The rest of the parameters are given in \cite{Caption1}. }
    \label{fig:isotropic}
\end{figure}

\subsection{Model Calculations}
Having established our general formalism, we next discuss the results of model-specific calculations of $\rho_{L}(\Omega)$ and $J_{c}(\Omega)$. In all cases, we describe the $C_{3z}$-symmetric part of the normal-state dispersion using a minimal model with nearest-neighbor dispersion at finite flux $\phi$ on the triangular lattice, 
\begin{align}
\label{eps_model}
    \epsilon_{\vec{k}} = - \sum_{j=1}^3 t \cos (\vec{a}_j\cdot \vec{k}-\phi/3).
\end{align} 
Here, $\vec{a}_j$ are three $C_{3z}$-related primitive vectors. In the following, we measure all energy scales in units of $t$, thus, setting $t\equiv 1$.

\subsubsection{$\vec{k}$-independent SC}
We start with a simple, $\vec{k}$-independent superconducting order parameter, which thus transforms under the trivial IR (with $d_{\Gamma}=1$) and the form factors are $f_{j}(\vec{k})=1$. As SC is isotropic, the rotational symmetry breaking must come from strain or nematic order in the normal state, described by the second term in \equref{NematicStrain}.
We see in \figref{fig:isotropic} that a strain/nematic field $\vec{\Phi}$ along the $x$ direction leads to maximum resistivity in $x$ as well, while the maximum  critical current is rotated by 90 degrees along $y$, leading to $\zeta_{xy} = -1$. The same conclusion was reached by the theoretical analysis of \cite{zhang2025angularinterplaynematicitysuperconductivity}, where, for a quadratic band structure, a Drude-like conductivity was shown to align with the direction of the highest critical current, contrary to the experimental observations. 

\subsubsection{Nematic SC}
To analyze nematic SC, we need to consider one of the two-dimensional IRs of the point group. For instance, for $D_6$ (or $C_6$ with displacement field), relevant to twisted bi- and trilayer graphene, these are $E_{1}, E_{2}$. The $E_{2}$ pairing is entirely band-off diagonal in presence of spin polarization \cite{BODPairingNatComm,PutzerBOD}, with many subtleties that are not of direct relevance here. We shall thus focus only on the $E_{1}$ representation with inter-valley pairing. The two possible sources of symmetry breaking for the nematic SC in a normal state can be classified into two broad categories:
\paragraph{Arising from $C_{3z}$ broken normal state.}
Here, the threefold rotational symmetry is already significantly broken in the normal electron liquid---either by a nematic transition or by a strain field---and this symmetry reduction is not the result of the SC.  
From this normal state, a superconducting instability emerges with order parameter transforming under $E_1$ of the (undistorted) point group with $C_{3z}$. Importantly, which order parameter direction $\vec{\eta} = (\eta_1,\eta_2)^T$ will be favored in the superconductor is determined by the normal-state anisotropy, linking normal-state and superconducting anisotropies.

\begin{figure}[tb]
    \centering
\includegraphics[width=1.0\linewidth]{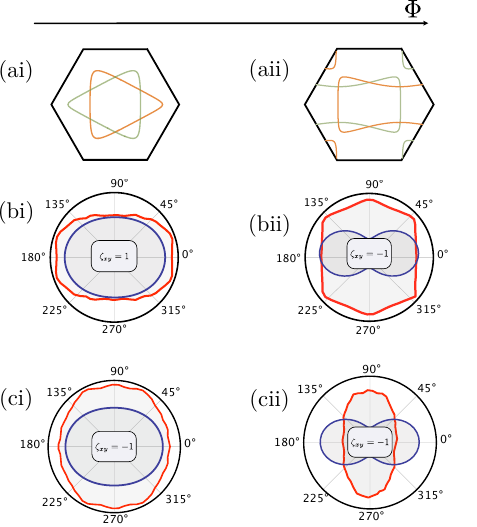}
    \caption{Nematic SC arising from $C_{3z}$ breaking in the normal state. (i) Weak nematic/strain order ($\Phi_{V} = 0.03$); (ii) strong nematic/strain order ($\Phi_{V} = 0.3$). (a) Fermi surfaces (FSs) for the two-valley model. The presence of nematic order along $x$ ($\theta = 0 $) breaks the $C_{3z}$ symmetry in the normal state. The critical currents (in red) and resistivity (in blue) are shown for (b) odd and (c) even form factors. Explicit parameters are same as  \cite{Caption1} except $\phi = -0.7\pi$.}
    \label{fig:nem_sc1}
\end{figure}

To study this, we investigate the superconducting part of the free-energy expansion, up to quadratic order given by \equref{eq:FreeEnergy} with $d_{\Gamma}=2$.
Assuming a strain field with $\vec{\Phi} = (\Phi_V,0)^T$ in the normal state, we first determine $\vec{\eta}$. The easiest way to do this is to compute the quadratic coefficients $a_{1}(\vec{q}=0)$ and $a_{2}(\vec{q}=0)$ and see which one is smaller. The expressions for the corresponding particle-particle bubbles are then given as \equref{ParticleParticleBubble}.
Although the exact point symmetries of the systems we consider here do not constrain $\vec{f}(\vec{k})$ to be even or odd in $\vec{k}$, we still discuss the even and odd cases separately. This can be thought of as limiting cases to probe the dependence of $\zeta_{xy}$ on $\vec{f}(\vec{k})$. We further note that twisted WSe$_2$ was found to be close to an emergent intravalley inversion symmetry with respect to the pairing instabilities, which can explain why the numerically obtained superconducting order parameters are indeed very close to being even or odd under $\vec{k} \rightarrow -\vec{k}$ \cite{7z4z-vlj8}. 
Let us first consider the case where $\vec{f}(\vec{k})$ is odd, $\vec{f}(\vec{k}) = -\vec{f}(-\vec{k})$. We calculate $a_{j}(\vec{q}=0)$ and find that $a_{1}(\vec{q}=0) < a_{2}(\vec{q}=0)$; as such, pairing with $\vec{\eta} = (\eta_1,0)^T$ and thus a superconducting order parameter $\Delta_{\vec{k}} \propto k_x$ is stabilized. 
Using the formalism described in \secref{sec:generalforma}, we next compute the angular dependence of resistivity $\rho(\Omega)$ and critical current $J_c(\Omega)$ (see \figref{fig:nem_sc1}). 
 The strain/nematic order parameter with $\vec{\Phi} = (\Phi_V,0)^T$, assumed to select the superconducting component as described above, implies $\rho_{xx} > \rho_{yy}$ and therefore the maximum of the resistivity aligns along $x$, irrespective of the value of $\Phi_V$. Interestingly, for small values of $\Phi_V$, we see that the maximum of the critical current is rotated (\figref{fig:nem_sc1} b(i)) compared to \figref{fig:isotropic} and is now larger along $x$ than $y$, leading to $\zeta_{xy}=1$. This is consistent with the observation of the experiment \cite{zhang2025angularinterplaynematicitysuperconductivity}. Increasing $\Phi$ further results in the case where the maximum  critical current is along $y$, which is perpendicular to the direction of maximum $\rho(\Omega)$, reverting to the naive expectation (\figref{fig:nem_sc1} b(ii)). 
 Intuitively, this is because at some point, the increase in the nematic order is so significant that the anisotropy coming from the superconducting order parameter is subdominant and the critical current asymmetry is the same as that of an isotropic superconducting order parameter, discussed in the previous subsection [cf.~\figref{fig:isotropic}(b)]. 

Performing the computation for the case where $\vec{f}(\vec{k}) \sim (2k_xk_y,k_y^2-k_x^2)$ is even, we see that for $\vec{\Phi}$ along $x$, again the first component is stabilized, i.e., $\Delta_{\vec{k}} \sim k_xk_y$. However, unlike the odd-$\vec{k}$ case, here the maximum critical current of the anisotropic superconductor for small normal-state anisotropy $\Phi_V$ is oriented already along the $y$ direction (see \figref{fig:nem_sc1} c(i)). As such, the directions of maximal resistivity and critical current are orthogonal here, which also does not change upon varying $\Phi_V$. 


\paragraph{Arising from vestigial nematic order.}
Next, we shall explore the scenario where the source of $C_{3z}$ symmetry breaking is a remnant nematic order persisting after the melting of a nematic superconducting order parameter---which itself arose from a $C_{3z}$-symmetric normal state at higher temperatures. As such, the reduced rotation symmetry in the superconducting and normal state are both rooted in superconducting correlations and, hence, again intertwined, although in a different way than in the previous scenario. 

To proceed, we adopt the action formalism, promoting the operators $c_{\vec{k},\nu}$ to Grassmann fields. As before, we focus on a superconducting order parameter, belonging to the $E_{1}$ representation, which explicitly breaks $C_{3z}$ rotational symmetry. The resulting action in Matsubara notation is given by $S=S_0 = S_{\text{int}}$, where $S_0 = \sum_{\nu}\int_{k} c_{k,\nu}^\dagger(\mathcal{G}^{-1}_{0})_{k,\nu}^{\pdagger} c^{\pdagger}_{k,\nu}$ and  
\begin{align}
\begin{split}
    S_{\text{int}}= & \int_{k,q} [\vec{f}(\vec{k})\cdot \vec{\eta}(q) c_{k+q,-}^{\dagger} c_{-k+q,+}^{\dagger}+\text{H.c.}]\\&+ \int_{q}\vec{\eta}^{\dagger}(q)\chi^{-1}(q)\vec{\eta}(q) + \int_{x} V(\vec{\eta}^{\dagger}(x),\vec{\eta}(x)). \label{FirstAction}
\end{split}
\end{align}
Here, $(\mathcal{G}_{0}^{-1})_{k,\nu} =(-i\omega_{n}+\xi_{\vec{k},\nu}) $ is the bare Green's function, with $k = (i\omega_{n},\vec{k})$ comprising  of 2D momentum and fermionic Matsubara frequencies; $\chi^{-1}(q)$ is the susceptibility taken to be of the form $\chi(q) = 1/(m^2 + \vec{q}^2 + \Omega_{n}^{2})$ where $q= (\vec{q},i\Omega_{n})$ comprises of the bosonic momenta and frequencies, and $V(\vec{\eta}^{\dagger},\vec{\eta})$ encodes the local bosonic interactions. We want $\vec{f}(C_{3z}\vec{k}) = C_{3z} \vec{f}(\vec{k})$ to describe pairing in the $E_1$ representation. Consequently, $V(\vec{\eta}^\dagger,\vec{\eta})$ has to be invariant under $\vec{\eta} \rightarrow C_{3z} \vec{\eta}$. Together with time-reversal symmetry, acting as $\vec{\eta} \rightarrow \vec{\eta}^*$, the only allowed (independent) terms are
\begin{equation}    V(\vec{\eta}^\dagger,\vec{\eta}) = u (\vec{\eta}^\dagger \cdot \vec{\eta})^2 + v (\vec{\eta}^\dagger \sigma_y \vec{\eta})^2 + \mathcal{O}(\eta^6), \label{Potential}
\end{equation}
in agreement with \cite{PhysRevResearch.2.033062}, where $\sigma_{j}$'s are the Pauli matrices. Negative $v$ favors a chiral pairing state, $\vec{\eta} = e^{i\varphi}(1,i)^T$ while $v > 0$ favors nematic pairing, $\vec{\eta} = (\cos \phi , \sin \phi)^T$. Note that the direction $\phi$ is only determined by higher-order terms \cite{PhysRevResearch.2.033062}, which we neglect here. We focus on $v>0$ from now on as we are interesting in nematic pairing.
The additional inclusion of the intra-valley in-plane rotation symmetry $C_{2y}$, acting as $c_{\vec{k},\nu} \rightarrow c_{(-k_x,k_y),\nu}$ on our fermions in \equref{FirstAction}, further requires that the basis functions obey $f_1(\vec{k})= -f_1(C_{2y}\vec{k})$ and $f_2(\vec{k})= f_2(C_{2y}\vec{k})$; meanwhile, the potential should then also be invariant under $\vec{\eta} \rightarrow \sigma_z \vec{\eta}$, which, however, does not further constrain \equref{Potential} up to quartic order---it only matters at higher orders.

To describe vestigial orders, we perform a Hubbard-Stratonovich decoupling \cite{PhysRevB.85.024534} of the quartic terms in $V$ such that \equref{FirstAction} effectively becomes
\begin{align}
\begin{split}
    S_{\text{int}} & \rightarrow \int_{k,q} [\vec{f}(\vec{k})\cdot \vec{\eta}(q) c_{k+q,-}^{\dagger} c_{-k+q,+}^{\dagger}+\text{H.c.}] \\& + \int_{q}\vec{\eta}^{\dagger}(q)\chi^{-1}(q)\vec{\eta}(q)+ \int_{x} \frac{1}{4v} \tilde{\vec{\Phi}}^2 + \int_{x} \frac{1}{4\tilde{u}} \Psi^2 \\& + \int_{x}\vec{\eta}^{\dagger}(i\Psi \sigma_0 + \tilde{\Phi}_{x}\sigma_{z} +\tilde{\Phi}_{y}\sigma_{x})\vec{\eta},
    \end{split} \label{SecondFormOfSint}
\end{align}
where $\tilde{u} = u+v$. We now discuss vestigial order in the large-$N$ limit of \cite{PhysRevB.85.024534}, where the saddle-point equations of the fields $\Psi$ and $\tilde{\Phi}_{x,y}$ become exact. As the saddle-point value $-i\Psi^0$ of $\Psi$ just leads to an isotropic mass renormalization of the bosons $\vec{\eta}$, we can neglect it here. Instead, we will focus on the saddle point value $\tilde{\vec{\Phi}}^0=(\tilde{\Phi}_x^0,\tilde{\Phi}_y^0)$ of $\tilde{\vec{\Phi}}$; the two components $\tilde{\Phi}_x^0$ and $\tilde{\Phi}_y^0$ correspond to the composite order parameters
\begin{equation}
    |\eta_1|^2 - |\eta_2|^2 \quad \text{and} \quad 2\, \text{Re} ( \eta_1^* \eta_2), \label{VestigialOP}
\end{equation}
respectively; these two components are gauge invariant, even under time-reversal, and transform under $E_2$. They hence describe the vestigial $C_{3z}$ symmetry breaking and represent the analogue of $\vec{\Phi}$ studied above. 
Symmetry dictates that if $\tilde{\vec{\Phi}}^{0}\neq0$, there will be three degenerate solutions of the saddle point equations, $\tilde{\vec{\Phi}}^0 = \tilde{\Phi}_0 \hat{\vec{n}}, \tilde{\Phi}_0 C_{3z}\hat{\vec{n}}, \tilde{\Phi}_0 C_{3z}^2\hat{\vec{n}}$ and the system will spontaneously choose one. Without $C_{2y}$, $\hat{\vec{n}}$ will point along a generic direction, while $C_{2y}$ will fix $\hat{\vec{n}}$ to be either $(1,0)^T$ or $(0,1)^T$.

For concreteness, we here assume that $\tilde{\Phi}_{y} \rightarrow 0$ and $\tilde{\Phi}_{x} \rightarrow \tilde{\Phi}_{x}^{0}$ upon condensation. This leads to distinct renormalized masses $m_{1}^{2} = m^2 + \tilde{\Phi}^{0}_{x}$ and $m_{2}^{2} = m^2 - \tilde{\Phi}^{0}_{x}$ for the two components of $\vec{\eta}$ and the associated action reduces to
\begin{align}
\begin{split}
        S_{\text{int}} &\rightarrow  \int_{k,q} [\vec{f}(\vec{k})\cdot \vec{\eta}(q) c_{k+q,-}^{\dagger} c_{-k+q,+}^{\dagger}+\text{H.c.}] \\&+ \int_{q}\vec{\eta}^{\dagger}(q)\begin{pmatrix}
m_{1}^2 + \vec{q}^2 + \Omega_{n}^{2} &0  \\
0 & m_{2}^2 + \vec{q}^2 + \Omega_{n}^{2}  \\
\end{pmatrix} \vec{\eta}(q), \label{VestigialDispTerm}
\end{split}
\end{align}
where we suppressed an irrelevant constant term. \\
{\it Mean-field theory---}We now integrate over the $\vec{\eta}(\vec{q})$ fields and perform a mean-field decoupling of the resultant effective electron-electron interaction (see \appref{app_intervalley}). Assuming that $\langle c_{\vec{k},+}^{\dagger}c_{\vec{k},-}^{\pdagger}\rangle = 0$, i.e., that there is no intervalley coherent (IVC) order or spontaneous TRS breaking and focusing only on the static $i\Omega_{n} = 0$ contributions, the resulting mean-field Hamiltonian describing a fluctuation-driven vestigial nematic order then reads  
\begin{equation}
    H^{\text{eff}}_{\text{MF}} = \sum_{\vec{k},\nu} \left(\epsilon_{\vec{k},\nu} -\sum_{j,\vec{q}}f_{j}^2(\vec{k})\frac{N_{\vec{k}-2\vec{q}}}{m_{j}^2 + \vec{q}^2 }\right)c_{\vec{k},\nu}^\dagger c_{\vec{k},\nu}^{\pdagger} \label{HeffMF}
\end{equation}
where $N_{\vec{k}}=\langle c_{-\vec{k},-}^{\dagger}c_{-\vec{k},-}^{\pdagger}\rangle = \langle c_{-\vec{k},+}^{\dagger}c_{-\vec{k},+}^{\pdagger}\rangle $. Equation~(\ref{HeffMF}) captures the contribution $\delta\epsilon_{\vec{k},\nu}$ in \equref{VestigialBrokenSyms}.
To demonstrate the features of this vestigial nematic phase explicitly, we now turn to explicit model calculations. For completeness and just as before, we consider cases with both odd and even form factors.  

\begin{figure}[tb]
    \centering
\includegraphics[width=\linewidth]{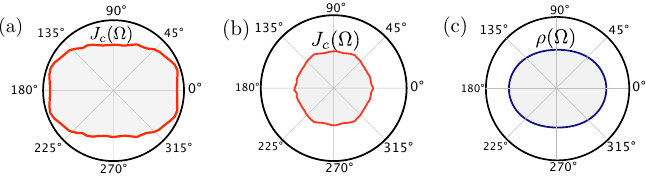}
    \caption{Critical current of a nematic superconductor with (a) odd form factor, (b) even form factor. Panel (c) depicts the angular dependence of resistivity arising from a vestigial nematic order. For both the even and odd cases, the angular dependence $\rho(\Omega)$ looks qualitatively identical. Parameters are the same as \cite{Caption1}.}
    \label{fig:nem_sc2}
\end{figure}

Assuming that the nematic state that is spontaneously selected is  $\vec{\eta}= (1,0)^{T}$, which means that we consider superconducting order parameters of the form $\Delta_{\vec{k}} \propto k_x$ (odd parity) or $\Delta_{\vec{k}} \propto k_x k_y$ for even parity. In either case, the resulting superconductor leads to the breaking of $C_{3z}$ symmetry, and consequently gives rise to an anisotropic critical current profile. Specifically, the critical current along the $x$ direction (see \figref{fig:nem_sc2}(a) for odd and (b) for even parity) exceeds that of $y$, leading to transport anisotropy characterized by $\text{sgn}(J_{c}(\Omega=0)-J_{c}(\Omega=\pi/2))=1$ for both cases, albeit with a weaker anisotropy in \figref{fig:nem_sc2}(b).  

When the superconducting order melts, the broken rotational symmetry associated with $\vec{\eta}= (1,0)^{T}$ remains encoded in $\tilde{\Phi}_x^{0}<0$ and $\tilde{\Phi}_y^{0} = 0$; this follows by noting that only the first component in \equref{VestigialOP} is non-zero and setting $\vec{\eta}= (1,0)^{T}$ in \equref{SecondFormOfSint} favors negative $\tilde{\Phi}_x$. Consequently, we have $m_1^2 < m_{2}^2$ in \equref{VestigialDispTerm}, which breaks the $C_{3z}$ symmetry and defines a vestigial nematic phase.
This defines $\delta\epsilon_{\vec{k},\nu}$ which enters \equref{FullExpressionForSigma} and allows us to obtain the result $\rho_{xx} > \rho_{yy}$, irrespective of the parity of the underlying form factor (see also as seen in \figref{fig:nem_sc2}(c)). 
Consequently, the direction of the maximum of the normal state resistivity from the nematic vestigial order parameter is aligned with the direction of the maximum critical current in the superconducting phase, leading to $\zeta_{xy}=1$.  This behavior is different from the previous case where the $C_{3z}$ breaking in the normal-state was unrelated to superconductivity and we found $\zeta_{xy}=-1$, except for odd-parity pairing and weak $\vec{\Phi}$, see \figref{fig:nem_sc1}.

\subsection{Nematic paraconductivity via Aslamazov-Larkin}
Instead of working with fermions within mean-field theory, we can work in an effective theory of dynamic bosonic order parameter fields $\vec \eta$, with dynamics endowed by integrating out fermions, and account for  coupling to vestigial fields $\tilde{\Phi}_{x/y}, \Psi$. Within this bosonic framework, one computes resistivity via the Aslamazov-Larkin diagram \cite{hecker_vestigial_2018}. This provides a complementary approach to mean-field theory, and we confirm here that the two approaches give consistent results---the relation $\rho_{xx}>\rho_{yy}$ for $\tilde{\Phi}_x^0<0$, $\tilde{\Phi}_y^0=0$, and $\Psi^0=0$.  

{\it Bosonic susceptibility---}Integrating out fermions, we find  from the couplings in \equref{FirstAction} the correction to the bosonic action given by
\begin{align}
\delta S_{\text{eff}}[\eta] &= \int_q\, \eta_i^*(\vec q)\, \delta\chi_{ij}^{-1}(\vec q)\, \eta_j(\vec q),\\
\notag \delta\chi^{-1}(\vec q)
&=\delta\chi_0^{-1}(\vec q)\sigma_0 + d_z (q_x^2-q_y^2)\sigma_z+ 2d_x q_x q_y \sigma_x.
\end{align}
Most important to the present analysis are the generated terms with prefactors $d_z,d_x$, which are not proportional to the identity matrix. 
By direct integration, taking $\epsilon_{\tau\vec k}$ of Eq. \eqref{eps_model} and sweeping across all fluxes $\phi$ and at a fixed temperature $T=0.2 t$, we find that $d_z<0$, $d_x>0$. 

The $\delta\chi_0^{-1}(\vec q)$ is absorbed into the diagonal component of the susceptibility. Including the off-diagonal components, we arrive at the bare susceptibility,
\begin{align}
\chi_0^{-1}&=(\Omega_n^2 + \vec q^2 + m^2)\sigma_0 + d_z(q_x^2-q_y^2) \sigma_z + 2 d_x  q_x q_y \sigma_x,
\end{align}
while the vestigial order-parameter correction to the susceptibility is,  given by
\begin{align}
\chi_\Phi^{-1}(\vec q)&=i\Psi^0\sigma_0+\tilde{\Phi}^0_x\sigma_z+\tilde{\Phi}^0_y\sigma_x,
\end{align}
which immediately follows from \equref{SecondFormOfSint},
upon taking the saddle point values $\Psi, \tilde{\Phi}_{x,y} \to \Psi^0, \tilde{\Phi}^0_{x,y}$. 
Armed with this, we now define the {\it full susceptibiliy} as
\begin{align}
 \chi^{-1}(\vec q)=\chi_0^{-1}(\vec q)+\chi_\Phi^{-1}(\vec q).
 \end{align}

 \begin{figure}[tb]
    \centering
\includegraphics[width=\linewidth]{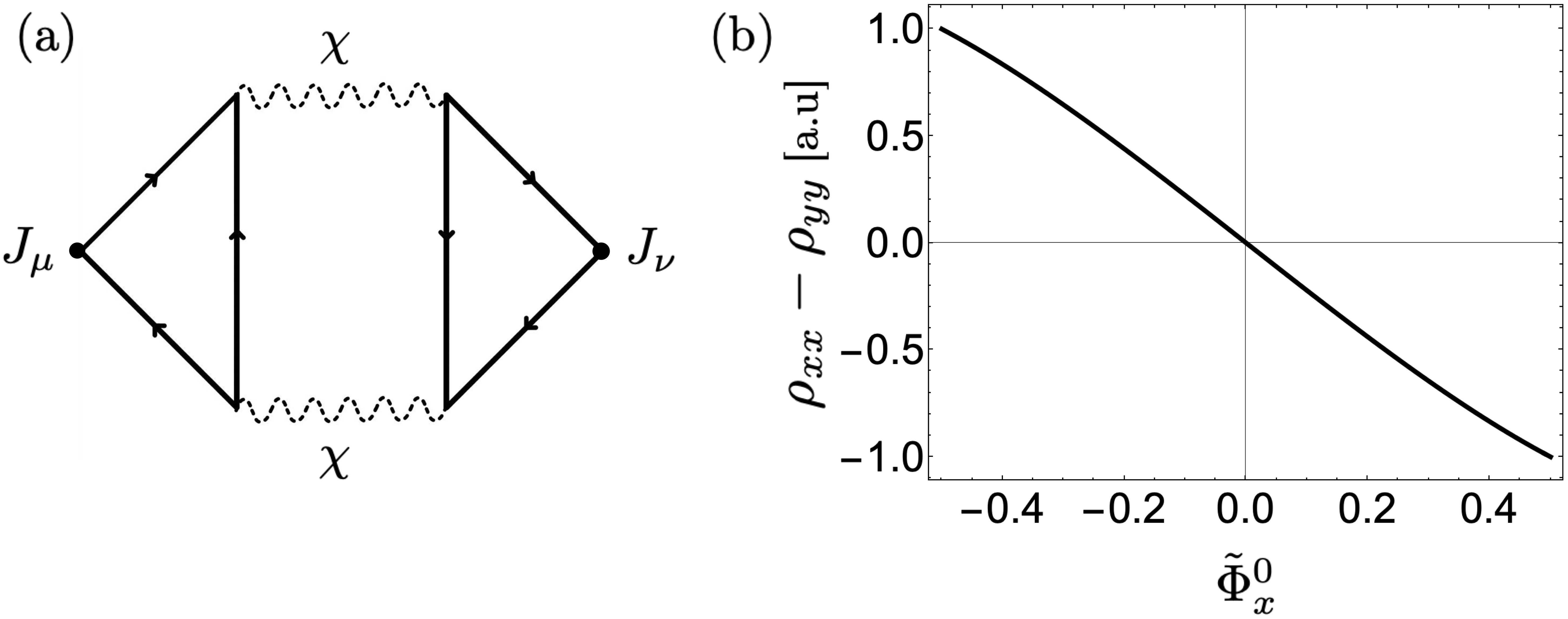}
    \caption{(a) Depiction of the Aslamazov-Larkin diagram. (b) Resistivity anisotropy, $\rho_{xx}-\rho_{yy}$ vs vestigial order $\tilde{\Phi}_x^0$, as computed via the Aslamazov-Larkin diagram.}
    \label{fig:AL}
\end{figure}

{\it Paraconductivity via Aslamazov-Larkin---}
The current vertex is obtained as $J_\mu=\partial_{q_\mu} \chi^{-1}(\vec q)$, i.e.
\begin{align}
\notag J_x&=2q_x\sigma_0 + 2d_z q_x \sigma_z+ 2d_x q_y \sigma_x,\\
J_y&=2q_y\sigma_0 + 2d_z q_y \sigma_z+ 2d_x q_x \sigma_x.
\end{align}
Meanwhile the conductivity tensor, as computed via the Aslamazov-Larkin diagram, is 
$\sigma_{\mu\nu}\propto \left(e^2/\hbar\right) \int_q \text{Tr} \left[J_\mu(\vec q) \chi(\vec q) \chi(\vec q) J_\nu(\vec q) \right]$ \cite{Aslamazov_1968a,Aslamazov_1968b}. Treating $d_{x/z}$ and $\tilde{\Phi}^0_{x,y}$ as expansion parameters, we obtain $\sigma_{xx}-\sigma_{yy} \propto  -d_z \tilde{\Phi}^0_x$ (and $\sigma_{xy} \propto  -d_x \tilde{\Phi}^0_y$). The anisotropy in the resistivity immediately follows, $\rho_{xx}-\rho_{yy} \propto  d_z \tilde{\Phi}^0_x$, see also \figref{fig:AL}(b) for the result obtained by direct computation of the diagram without expansion. Finally, since $d_z<0$, then $\sign[\rho_{xx}-\rho_{yy} ] = -\sign(\tilde{\Phi}^0_x)$, which is consistent with our mean-field treatment of the vestigial phase in the previous subsection.

\section{Intravalley pairing}\label{Intravalley}
With observations of pairing in rhombohedral tetralayer graphene and twisted MoTe$_{2}$ \cite{han_signatures_2025,MoTe2_unconv_exp} emerging from a valley-imbalanced normal state as our inspiration, we next examine an analogous scenario in a model with only one active valley degree of freedom and, hence, broken time-reversal symmetry in the normal state. As at least in some of those superconducting phases also spin seems to be polarized, we further start from spinless electrons with creation operators $c_{\vec{k}}^\dagger$. Denoting the single-valley dispersion by $\epsilon_{\vec{k}}$, the Hamiltonian in this case reads as
\begin{equation}
    \mathcal{H} = \sum_{\vec{k}} c^\dagger_{\vec{k}} c^\pdagger_{\vec{k}} \epsilon_{\vec{k}}^{\pdagger} + \sum_{\vec{k},\vec{q}} \left[\Delta_{\vec{q}} f(\vec{k})c^\dagger_{\vec{k}+\vec{q}} c^\dagger_{-\vec{k}+\vec{q}}  + \text{H.c.}\right], \label{SingleValleySC}
\end{equation}
where $f(\vec{k})$ encodes the nature of the pairing state. Since it must obey $f(\vec{k}) = - f(-\vec{k})$ as a result of Fermi-Dirac statistics, we choose a chiral superconducting order parameter with $f(\vec{k}) \sim k_x \pm i k_y$ (for small $\vec{k}$; appropriately extended to the full Brillouin zone), which will lead to a full gap and preserve $C_{3z}$ symmetry in gauge-invariant observables. In fact, this type of chiral state is expected to be favored, at least within mean-field theory, in line with many theoretical works \cite{rhomohedral1,yoon2025quartermetalsuperconductivity,maymann2025pairingmechanismdictatestopology,wy3f-hgr9,geier2024chiraltopologicalsuperconductivityisospin,yang2024topologicalincommensuratefuldeferrelllarkinovchinnikovsuperconductor,qin2025chiralfinitemomentumsuperconductivitytetralayer,jahin2025enhancedkohnluttingertopologicalsuperconductivity,2025arXiv250305697M,christos2025finitemomentumpairingsuperlatticesuperconductivity,sedov2025probingsuperconductivitytunnelingspectroscopy,gil2025chargepairdensitywaves,zfmh-rjzc,k8sb-rqxf,2024arXiv241108969S}, with the sign $\pm$ determined by the chirality of the normal state \cite{2025arXiv250305697M,christos2025finitemomentumpairingsuperlatticesuperconductivity}. For concreteness, we will focus on $f(\vec{k}) \sim k_x + i k_y$ in the following.

Importantly, as a result of the broken time-reversal symmetry associated with valley imbalance, the maximum of the particle-particle $\Gamma(\vec{q})$ might not be at $\vec{q}=0$ \cite{scammell_theory_2022}, leading to finite-momentum pairing already in equilibrium; this is the regime we will focus on here. In the presence of $C_{3z}$ in the normal state, there will be three degenerate maxima at positions $\vec{q}_j = (C_{3z})^{j-1} \vec{q}_1$, $j=1,2,3$. To define a free-energy expansion, we allow for a superposition of all three momenta in \equref{SingleValleySC} and write $\Delta_{\vec{q}} = \sum_{j=1}^3 \eta_j(\delta \vec{q}) \delta_{\vec{q},\vec{q}_j+\delta\vec{q}}$, where $\delta\vec{q}$ is a small momentum deviation from the minima, which will be used to parametrize fluctuations around the equilibrium configuration and to impose a finite supercurrent later.  
Up to quartic order in $\eta_j$, the free energy reads as
\begin{align}
        \mathcal{F}_{3\vec{q}} &\sim \sum_{\delta\vec{q},j=1}^{3} a_{\vec{q}_{j}+\delta \vec{q}}|\eta_j(\delta \vec{q})|^{2} + \int_x V_{3\vec{q}}(\eta_j^*(\vec{x}),\eta_j(\vec{x})),
\end{align}
where we wrote the quartic terms, with 
\begin{equation}
    V_{3\vec{q}}(\eta_j^*,\eta_j) = u\left(\sum_{j=1}^{3}|\eta_j|^{2}\right)^2 + v \sum_{j\neq j'=1}^{3}|\eta_j|^{2}  |\eta_{j'}|^{2}, \label{InteractionPotential}
\end{equation}
in real space---for future reference and to keep the notation more compact. Here, $\eta_j(\vec{x})$ is the Fourier transform of $\eta_j(\delta \vec{q})$.
Depending on the sign of $v$, we either obtain a ``$1\vec{q}$ state'' or a ``$3\vec{q}$ state''. In the case of the former, only one of $\eta_{1,2,3}$ is non-zero such that translational symmetry is preserved in observables, while $C_{3z}$ is broken. In the $3\vec{q}$ state, $C_{3z}$ is preserved while translational invariance is broken \cite{sedov2025probingsuperconductivitytunnelingspectroscopy}.

It was recently observed in \cite{morissette2025coulombdrivenmomentumspacecondensation} that at the onset of valley polarization in rhombohedral graphene also a nematic transition occurs. Upon entering the superconducting phase from this nematic phase, another experiment \cite{morissette2025stripedsuperconductorrhombohedralhexalayer}  provides evidence for ``striped'' superconductivity. 
The plethora of intertwined phases with different sorts of symmetry breaking (translational, time-reversal) in such superconductors naturally motivates the question regarding the emergence of vestigial phases, and how these phases might influence the directional dependence on transport properties. This is the question we turn to next. To this end, we first devise a theoretical framework for the corresponding vestigial orders.

\subsection{Vestigial phases of superconducting order}
To develop a theory of vestigial phases of this form of finite-momentum superconductivity, we start from the associated action description, 
\begin{align}\begin{split}
    S_{3q} &= S_0 + \sum_{j=1}^3 \int_{k, \delta q} \eta_j(\delta q) f(\vec{k}) c_{k+q_j+\delta q}^{\dagger} c_{-k+q_j+\delta q}^{\dagger} \\
    &
 \quad + \sum_{j=1}^3 \int_{\delta q}\left|\eta_j(\delta \vec{q})\right|^2\left(m^2+\delta \vec{q}^2+\Omega_n^2\right) + S_{V}, 
\end{split}\end{align}
where $S_0= \int_{k} c_{k}^\dagger(-i\omega_{n}+\epsilon_{\vec{k}}) c^{\pdagger}_{k}$ and
\begin{equation}
    S_{V} =  \int_{x}\left[ u (\text{tr}[\rho])^2+ v \sum_{j\neq j'=1}^{3} \rho_{j,j'}\rho_{j',j}\right].
\end{equation}
Here, we introduced $\rho_{j,j'}= \eta_{j}^{*}\eta_{j'}$ as possible composite order parameters. As before, we decouple the quartic terms using Hubbard-Stratonovich fields $\psi_{j,j'}$ such that effectively
\begin{align}
\begin{split}
S_{V} \rightarrow &  \int_{x}\left[\sum_{j\neq j'}[\rho_{j,j'}\psi_{j',j} + \text{H.c}] +i\sum_{j}\phi_{j}\rho_{j,j}\right] \\&+\int_{x}\sum_{j}\frac{1}{4u}\phi^2_{j}  - \int_{x}\sum_{j\neq j'}\frac{1}{2\tilde{v}}\psi_{j,j'}\psi_{j',j}.
       \end{split}
\end{align}
where $\tilde{v} = u +v$. We then treat $\psi_{j,j'},\phi_{j}$ on the saddle-point level, which becomes exact in the limit where the number of components of $\vec{\eta}$ is infinitely large (large $N$ limit). For instance, for the $1\vec{q}$ state, one would only have one of the components of $\phi_{j}$ non-zero, i.e., $i\phi_{j} = \psi_0\delta_{j,1}$, without loss of generality. But other configurations are also possible. Driven by the above-mentioned experimental claim of striped superconductivity \cite{morissette2025stripedsuperconductorrhombohedralhexalayer} and the fact that this order leads to particularly interesting symmetry breaking, we here focus on a $2\vec{q}$-state (and defer the discussion of the $1\vec{q}$ state to \appref{app_1q}). On the level of the saddle-point fields, the associated vestigial phase is characterized by $\psi_{2,3} = \psi_{3,2}^* =  \psi_{0} \neq 0$ (picking $2$ and $3$ without loss of generality) with all other components zero.
Integrating out the $\vec{\eta}$ fields, yields an effective $2\vec{q}$ action (see \appref{app_intravalley}) $S^{\text{eff}}_{2q} = S_0 + S_{1} + S_{2}$ with
\begin{subequations}\begin{align}
        S_{\text{1}} &= -\sum_{j}\int_{\delta q} \frac{1}{[\tilde{\chi}_{j}(\delta q)]^{-1}} D_{j}^{\dagger}(\delta \vec{q})D_{j}^{\pdagger}(\delta \vec{q}),\\
        S_{2} &= \int_{\delta q} \frac{1}{[M_{j=2}(\delta q)]^{-1}}(\psi_0 D_2^{\dagger}(\delta \vec{q}) D_3^{\pdagger}(\delta \vec{q})
        +\text{H.c.}).
\end{align}\end{subequations}
Here $D_{j}(\delta \vec{q}) = \sum_{k}^{\pdagger} f(\vec{k}) c_{-k+q_j+\delta q}^{\pdagger} c_{k+q_j+\delta q}^{\pdagger}$ is a fermionic bilinear and the renormalized propagators are defined by $[M_{j}(\delta q)]^{-1} =[\chi(\delta q)]^{-2}-\left|\psi_0\right|^{2}(\delta_{j,2}+\delta_{j,3})$ and $[\tilde{\chi}_{j}(\delta q)]^{-1} = [M_{j}(\delta q)]^{-1}[\chi(\delta q)]$. As expected, the $j=1$ component remains unchanged, whereas the propagators for $j=2,3$ are renormalized due to the presence of the $\psi_{0}$ field. Notably, as a consequence of the off-diagonal nature of $\psi_{0}$, the term $S_{2}$ explicitly breaks translational symmetry through a non-zero momentum transfer vector $2\vec{Q}$ where $\vec{Q} = \vec{q}_{3}-\vec{q}_{2}$. As a result, two distinct symmetries are broken: $C_{3z}$ rotational symmetry associated with nematic order, and translational symmetry along the direction of $\vec{Q}$. As such, we obtain a uni-directional and, thus, nematic vestigial charge-density wave phase emerging out of the superconductor.

{\it Mean-field theory---}To be able to compute the associated transport anisotropies, we subsequently go back to a Hamiltonian description and perform a mean-field decoupling, assuming no additional symmetries are broken. As detailed in \appref{app_intravalley}, we finally arrive at an effective Hamiltonian of the form:
\begin{equation}
    \mathcal{H}_{\text{MF}} = \sum_{\vec{k}}[\epsilon_{\vec{k}}-\tilde{\alpha}(\vec{k})] c_{\vec{k}}^{\dagger}c_{\vec{k}}^{\pdagger} +  \beta \sum_{\vec{k}}c^\dagger_{\vec{k}+2\vec{Q}}c_{\vec{k}}^{\pdagger}. \label{MFHamCDW}
\end{equation}
Here $\tilde{\alpha}(\vec{k}) = \sum_{j}\Lambda_{j}|f(\vec{k}-\vec{q}_{j})|^{2}n_{\text{F}}(\epsilon_{\vec{k}})$, where $\Lambda_{j} = \Lambda(\delta_{j,2} + \delta_{j,3})$ and $\beta$ are mean-field parameters. 
The first correction $\propto \tilde{\alpha}(\vec{k})$ in \equref{MFHamCDW} corresponds to a translational symmetry-obeying yet in general nematic contribution. The second correction ($\propto \beta$) is a translational symmetry-breaking term, as indicated by the coupling of momenta related by $2\vec{Q}$.

\begin{figure}[tb]
    \centering
\includegraphics[width=\linewidth]{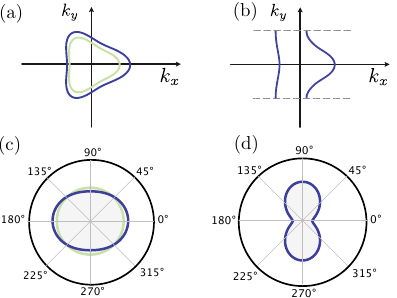}
    \caption{Normal state vestigial phases and their resistivities. In (a,c) we show the Fermi  surfaces and resistivities for the cases with $\beta=0$, and in (b,d) with $\tilde{\alpha}(\vec{k})=0,\beta\neq 0$. In (a,c) the green FS and resistivity curve indicates the case without nematicity, whereas the dark blue curves are with $\Lambda_{1}= 0, \Lambda_{2}=\Lambda_{3}=\Lambda\neq 0$. The Brillouin-zone strip is indicated by the dashed lines in (b). }
\label{fig:2q_FS}
\end{figure}

In \figref{fig:2q_FS}, we illustrate the resultant FS of the associated vestigial orders. For $\epsilon_{\vec{k}}$ we here use a continuum model with a quadratic dispersion, deformed by a trigonal warping term $\gamma$, i.e., $\epsilon_{\vec{k}} = \vec{k}^2[1+ \gamma \cos(3\tilde{\theta}_{\vec{k}})] - \mu$.  
With both $\tilde{\alpha}(\vec{k}),\beta = 0$, the FS retains its expected $C_{3z}$ with trigonal warping (shown in green in \figref{fig:2q_FS}(a)) which leads to an isotropic resistivity. When $\tilde{\alpha}(\vec{k})$ is now turned on---on account of the vestigial phase of the parent $2\vec{q}$ superconductor with $\psi_{2,3}\neq 0$, the parameters satisfy $\Lambda_{1} =0, \Lambda_{2}=\Lambda_{3} \neq 0$. This leads to breaking of the three-fold rotational symmetry, resulting in an anisotropic FS as can be seen in dark blue in \figref{fig:2q_FS}(a). The associated elongation along $x$ induces a resistivity maximum along $x$, as can be seen in \figref{fig:2q_FS}(c).

To isolate the impact of the term $\propto \beta$ in \equref{MFHamCDW} and compare with the previous case, we next consider $\tilde{\alpha}(\vec{k})=0$ but $\beta\neq 0$. This leads to charge-density modulations, where the band copies separated by the wave vector $2\vec{Q}$ hybridize. With our $\vec{Q}$ along the $y$ direction, the Brillouin zone is finite along $k_y$ but still infinite along $k_x$. This can be seen in \figref{fig:2q_FS}(b) where we find open FSs resulting from bands with much smaller velocities along $y$ than along $x$.
This is also why the current flow is much more impeded along $y$ compared to $x$, leading to a resistivity maximum in that direction. This is in sharp contrast to the case where $\tilde{\alpha}(\vec{k})\neq0,\beta=0$ where the associated orientation of resistivity is orthogonal to this case. Consequently, the competition between the translational-symmetry breaking ($\beta$) and nematic contribution ($\alpha$) in \equref{MFHamCDW} determines the sign of $\rho_{L}(\Omega=0)-\rho_{L}(\Omega=\pi/2)$.

\subsection{Critical current of $2\vec{q}$ state}
\begin{figure}[tb]
    \centering
\includegraphics[width=\linewidth]{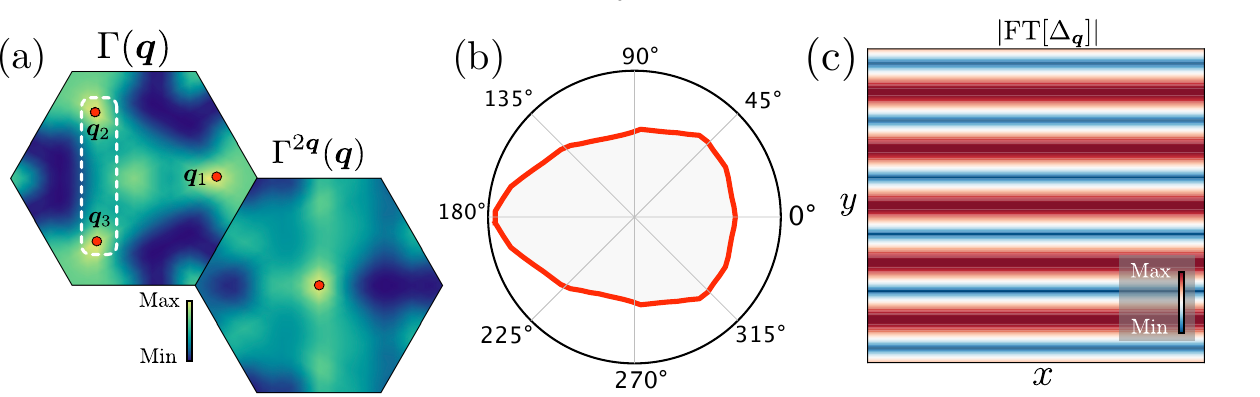}
    \caption{2$\vec{q}$ superconductor. (a) shows $\Gamma(\vec{q})$ and $\Gamma^{2\vec{q}}(\vec{q})$ where the red-dots indicate the maxima of the particle particle bubbles; (b) shows the angular dependence of critical current which is seen to break $C_{3z}$ symmetry as expected. Here $\phi = -0.8 \pi$, the rest of the parameters remain the same \cite{Caption1} and panel (c) shows real space modulation of Fourier transform (FT) of $|\Delta_{\vec{q}}|$, illustrating stripe order.}
\label{fig:3q_Currentprofile}
\end{figure}

Having established the normal-state transport anisotropy, we next turn to the parent superconductor where not only the composite order parameter $\rho_{2,3} = \eta^*_2\eta_3^{\phantom{*}}$ is condensed, but also $\eta_2$ and $\eta_3$ individually. With only two of the three components $\eta_{1,2,3}$ non-zero, rotation and translation symmetries are, of course, also broken in the superconductor and we have a ``striped superconductor''. 
In order to compute its critical currents, we adopt a minimal setting, where the two stabilized components $\eta_2$ and $\eta_3$ are assumed to continue to have equal values, $\eta_2(\delta\vec{q})=\eta_3(\delta\vec{q})=:\eta_{2q}(\delta\vec{q})$, even when we de-tune their momentum from their respective equilibrium values ($\delta\vec{q} \neq 0$). Then, the free energy can be effectively written as
\begin{equation}
    \mathcal{F}_{2\vec{q}} = \sum_{\delta \vec{q}}[a_{\vec{q}_{2}+\delta\vec{q}} +a_{\vec{q}_{3}+\delta\vec{q}} ] |\eta_{2q}(\delta\vec{q})|^{2} + \mathcal{O}(\Delta^4).  \label{2qGLExpansion}
\end{equation}
Our procedure to construct the particle-particle bubble for the $2\vec{q}$ state is as follows. First, we compute the particle-particle bubble [see \equref{eq:pp_intravalley} for expression] to determine $a_{\vec{q}}$ in \equref{2qGLExpansion} and the three $\vec{q}_{j}$ associated with superconducting instability. Then, to model the $2\vec{q}$ state, we restrict ourselves to retain $\vec{q}_{2}$ and its $C_{3z}$-rotated partner $\vec{q}_{3}$, highlighted with white dashed lines in \figref{fig:3q_Currentprofile}(a), upper panel. The resulting effective particle-particle bubble of the $2\vec{q}$ state from \equref{2qGLExpansion} is then given by
\begin{equation}
    \Gamma^{2\vec{q}}(\delta \vec{q}) = \Gamma(\vec{q}_{2}+\delta \vec{q}) + \Gamma(\vec{q}_{3}+\delta \vec{q}).
\end{equation}
In \figref{fig:3q_Currentprofile} (a) we show $\Gamma(\vec{q})$ and the corresponding $\Gamma^{2\vec{q}}(\vec{q})$, obtained from the above procedure, in the upper and lower panel, respectively. By design, $\Gamma^{2\vec{q}}(\vec{q})$ now has a maximum at $\vec{q}=0$ and explicitly breaks $C_{3z}$ symmetry. As before,  the current is given by$\vec{J}^{2\vec{q}}(\vec{q}) = 2 e \Gamma^{2\vec{q}}\vec{\nabla}_{\vec{q}}\Gamma^{2\vec{q}}$ from which we extract the critical currents. As expected, since both time reversal and inversion symmetry are broken, the critical currents in opposite directions are unequal, leading to a non-zero diode effect \cite{han_signatures_2025,MoTe2_unconv_exp,scammell_theory_2022, zgnk-rw1p,PhysRevLett.132.046003,2025arXiv250217555Y}. However,  this effect is not the focus of our work, since we are primarily interested in (sources of) rotational symmetry breaking, and will not pursue the diode effect further. 

The resulting angular dependence of the critical current is shown in \figref{fig:3q_Currentprofile}(b). The rotational symmetry is clearly broken, with the extrema of the critical current pointing in the direction $\pm \hat{\vec{q}}_{1}$, i.e., the momentum component that remains uncondensed. This behaviour is consistent with the real-space modulation of the superconductor (see \figref{fig:3q_Currentprofile}(c)), which exhibits stripe order oriented in the $y$ direction. 

Comparing with the angular dependency of the resistivity of the nematic order $\zeta_{xy} =1$ while for the charge density wave order $\zeta_{xy} = -1$. This
is consistent with the observations in \cite{morissette2025stripedsuperconductorrhombohedralhexalayer}, where it was also noted that in the SC displaying stripe order, the maximum of critical current is anti-correlated with that the direction of maximum resistivity in the metallic state above the superconducting transition temperature.  

\section{Conclusion}
In this work, we have studied the relation between the angular dependence of the normal-state resistivity $\rho(\Omega)$ and of the critical current $J_c(\Omega)$ for different  microscopic origins of rotational symmetry breaking, which we summarize in \tableref{SummaryResults}. Our study is primarily motivated by two-dimensional systems, such as rhombohedral graphene or twisted van der Waals materials, where these angular dependencies can be accessed via sunbeam Hall bar geometries \cite{chichinadze2025observationgiantnonlinearhall,morissette2025stripedsuperconductorrhombohedralhexalayer} and rotational symmetry breaking and superconductivity are ubiquitous in the phase diagram. This is why we use models with two valley degrees of freedom and a hexagonal lattice symmetry, although the formalism can be straightforwardly applied to other scenarios as well.

We started with the scenario in which the only source of rotational symmetry breaking comes from the normal state, while the superconducting order parameter is isotropic. In that case, the maxima of the critical current and of the normal-state resistivity occur along perpendicular directions. Since these maxima need not be perfectly aligned with high-symmetry directions of the lattice, we here use $\zeta_{xy} = -1$ [see \equref{DefinitionOfZeta} for formal definition] to compactly describe the situation of misaligned anisotropies along the two perpendicular high-symmetry directions $\Omega = 0$ and $\pi/2$. 
We then contrasted this behavior with the scenario where also the superconducting order itself is nematic, i.e., transforms non-trivially under rotations, taking into account that the normal-state anisotropies determine the specific superconducting order parameter configuration. Here, the result depends on the parity of the superconducting order parameter $\Delta(\vec{k})$. If $\Delta(\vec{k}) = \Delta(-\vec{k})$ (``nematic, even'' in \tableref{SummaryResults}), we obtain the same behavior as in the first scenario with an isotropic pairing state. However, an odd-parity order parameter, $\Delta(\vec{k}) = -\Delta(-\vec{k})$, yields the opposite behavior, $\zeta_{xy} = 1$, if the strain/nematic order in the normal state is sufficiently weak. If the latter is large enough to dominate the anisotropy inside the superconductor, we are then effectively again back to the first scenario and find $\zeta_{xy} = -1$.

The third distinct type of scenario arises when the superconductor is the \textit{only} source of broken rotational symmetry, including in the resistive state via the emergence of a vestigial nematic state. We show that this leads to the opposite behavior, $\zeta_{xy} = 1$, i.e., the maximal critical current and normal-state resistivity are along the same direction, irrespective of the parity of the order parameter.

\begin{table}[tb]
\begin{center}
\caption{Summary of our results, where $\parallel$ ($\perp$) indicates that the direction of maximum critical current of the superconductor is parallel (perpendicular) to the direction of maximum resistance in the normal state, or, more precisely, $\zeta_{xy} = 1$ ($\zeta_{xy} = -1$) with $\zeta_{xy}$ as defined in \equref{DefinitionOfZeta}. The arrow in the second line signals how the behavior changes upon increasing the strength of the anisotropy in the normal state. The normal state has time-reversal symmetry in all cases, except for the last two lines where we take it to be valley polarized leading to finite center-of-mass pairing $\vec{q}\neq 0$.}
\label{SummaryResults}\begin{ruledtabular}\begin{tabular}{ccc}
Superconductor & Normal State & Orientation   \\ \hline
isotropic & strain/other nematic & $\perp$ \\
nematic, odd & strain/other nematic & $\parallel$ $\rightarrow$  $\perp$ \\
nematic, even & strain/other nematic & $\perp$ \\ \hline
nematic, odd/even & vestigial nematic & $\parallel$ \\
$2\vec{q}$ PDW & \begin{tabular}[c]{@{}c@{}}vestigial CDW dominant \\ vestigial nematic dominant\end{tabular} & \begin{tabular}[c]{@{}c@{}} $\perp$ \\$\parallel$  \end{tabular}
 \end{tabular}
\end{ruledtabular}
\end{center}
\end{table}

Finally, we also considered pairing within a single valley, i.e., when time-reversal symmetry is already broken in the normal state, which naturally leads to pairing at finite center-of-mass momentum $\vec{q} \neq 0$. This, in turn, allows for interesting forms of non-superconducting vestigial orders. Motivated by the experimental results in \refcite{morissette2025stripedsuperconductorrhombohedralhexalayer}, we focus on a translational-symmetry-breaking unidirectional pair-density wave ($2\vec{q}$ PDW in \tableref{SummaryResults}) which also leads to a charge-density modulated vestigial phase in the normal state. We show that both $\zeta_{xy} = -1$ and $\zeta_{xy} = 1$ are possible, depending on whether the charge-density modulations or additional translational-symmetry-preserving terms associated with the superconducting vestiges dominate. 

Taken together, our findings show that these relative transport anisotropies in the superconductor and related normal state can provide crucial insights. For instance, the maximum critical current and normal-state resistance being aligned, as seen in \cite{zhang2025angularinterplaynematicitysuperconductivity}, indicates that the superconducting order parameter is either odd parity with its direction determined by a normal-state anisotropy that is unrelated to the superconductor or the dominant source of rotational symmetry breaking comes from the superconductor alone, including in the normal state due to nematic paraconductivity or vestigial nematic order. If the relative anisotropy $\zeta_{xy}$ is found to change sign, e.g., as a function of displacement field or filling, only the former of the two scenarios is naturally consistent. Furthermore, our findings for the valley polarized case show that the observations of \refcite{morissette2025stripedsuperconductorrhombohedralhexalayer}, where $\zeta_{xy} = -1$, imply that the charge-density modulated part of the vestigial order dominates.

\begin{acknowledgments}
S.B.~and M.S.S.~acknowledge funding by the European Union (ERC-2021-STG, Project 101040651---SuperCorr). Views and opinions expressed are however those of the authors only and do not necessarily reflect those of the European Union or the European Research Council Executive Agency. Neither the European Union nor the granting authority can be held responsible for them. M.S.S.~thanks J.~Li for bringing the question of correlations between normal-state and superconducting transport anisotropies to our attention. 
\end{acknowledgments}
\bibliography{bib_nematic}

\onecolumngrid

\begin{appendix}

\section{Intervalley pairing: Mean field theory of nematic vestigial order}\label{app_intervalley}
In this section, we derive the mean-field theory for the vestigial order arising from the nematic superconductor with pairing in between valleys. Reiterating the effective action from the main text,
\begin{align}
\begin{split}
        S_{\text{int}} &=  \int_{k,q} [\vec{f}(\vec{k})\cdot \vec{\eta}(q) c_{k+q,-}^{\dagger} c_{-k+q,+}^{\dagger}+\text{H.c.}]+ \int_{x} \frac{1}{4v} \vec{\tilde{\Phi}}^2 + \int_{q}\vec{\eta}^{\dagger}(q)\begin{pmatrix}
m_{1}^2 + \vec{q}^2 + \Omega_{n}^{2} &0  \\
0 & m_{2}^2 + \vec{q}^2 + \Omega_{n}^{2}  \\
\end{pmatrix} \vec{\eta}(q).
\end{split}
\end{align}
We now integrate out the fields $\vec{\eta},\vec{\eta}^{\dagger}$ to obtain a four-fermion effective interaction of the form
\begin{equation}
    S_{\text{int}}^{\text{eff}}=- \sum_{k,k',q,j} \frac{f_{j}(\vec{k})f_{j}(\vec{k}')}{m_{i}^2 + \vec{q}^2 + \Omega^2_n}c_{k+q,-}^{\dagger}c_{-k+q,+}^{\dagger}c_{-k'+q,+}^{\pdagger}c_{k'+q,-}^{\pdagger}.
\end{equation}

From this, we then obtain an effective Hamiltonian of the form 
\begin{equation}
    H^{\text{eff}} = \sum_{\vec{k},\nu} \xi_{\vec{k},\nu} c_{\vec{k},\nu}^\dagger c_{\vec{k},\nu}^{\pdagger} - \sum_{\vec{k},\vec{k'},\vec{q},j=1,2} \frac{f_{j}(\vec{k})f_{j}(\vec{k}')}{m_{j}^2 + \vec{q}^2 }c_{\vec{k}+\vec{q},-}^{\dagger}c_{-\vec{k}+\vec{q},+}^{\dagger}c_{-\vec{k}'+\vec{q},+}^{\pdagger}c_{\vec{k}'+\vec{q},-}^{\pdagger}
\end{equation}
Performing a mean-field decoupling in the direct channel, we get
\begin{equation}
      H^{\text{eff}}_{\text{MF}} = \sum_{\vec{k},\nu} \epsilon_{\vec{k},\nu} c_{\vec{k},\nu}^\dagger c_{\vec{k},\nu}^{\pdagger} - \sum_{\vec{k},\vec{q},j=1,2} \frac{f_{j}^2(\vec{k})}{m_{j}^2 + \vec{q}^2 }[\langle c_{\vec{k}+\vec{q},-}^{\dagger}c_{\vec{k}+\vec{q},-}^{\pdagger}\rangle c_{-\vec{k}+\vec{q},+}^{\dagger}c_{-\vec{k}+\vec{q},+}^{\pdagger}+ c_{\vec{k}+\vec{q},-}^{\dagger}c_{\vec{k}+\vec{q},-}^{\pdagger}\langle c_{-\vec{k}+\vec{q},+}^{\dagger}c_{-\vec{k}+\vec{q},+}^{\pdagger}\rangle]
\end{equation}
With $\langle c_{-\vec{k},-}^{\dagger}c_{-\vec{k},-}^{\pdagger}\rangle = \langle c_{-\vec{k},+}^{\dagger}c_{-\vec{k},+}^{\pdagger}\rangle =N_{\vec{k}}$, we get
\begin{equation}
    H^{\text{eff}}_{\text{MF}} = \sum_{\vec{k},\nu} \left(\epsilon_{\vec{k},\nu} -\sum_{j,\vec{q}}f_{j}^2(\vec{k})\frac{N_{\vec{k}-2\vec{q}}}{m_{j}^2 + \vec{q}^2 }\right)c_{\vec{k},\nu}^\dagger c_{\vec{k},\nu}^{\pdagger}.
\end{equation}
We take $N_{\vec{k}} \sim n_{\text{F}}(\epsilon_{\vec{k}})$ where $n_{\text{F}}(.)$ is the Fermi distribution function as a first order approximation (``one shot Hartree Fock") to self-consistent mean-field solution $N_{\vec{k}}$. 

\section{Critical currents in $1\vec{q}$ states}\label{app_1q}
In this section, we outline the computation of critical currents for intra-valley pairing.
Neglecting the 3$\vec{q}$ state for now, we will assume that $\Delta_{\vec{q}}$ is only non-zero for a single $\vec{q}$ at a time. Importantly, since TRS is already broken, this equilibrium pairing momentum $\vec{q}$ can itself be non-zero.

Using the triangular lattice dispersion $\epsilon_{\vec{k}}$ in \equref{eps_model}, and performing the same steps as before, one obtains the particle-particle bubble as
\begin{equation}
    \Gamma(\vec{q})  = \frac{1}{2N}\sum_{\vec{k}} \frac{\tanh{\frac{\epsilon_{\vec{k}+\vec{q}}}{2 T}} + \tanh{\frac{\epsilon_{-\vec{k}+\vec{q}}}{2 T}}}{\epsilon_{\vec{k}+\vec{q}} + \epsilon_{-\vec{k}+\vec{q}}}|f(\vec{k})|^2.
    \label{eq:pp_intravalley}
\end{equation}

\begin{figure}[h]
    \centering
\includegraphics[width=0.5\linewidth]{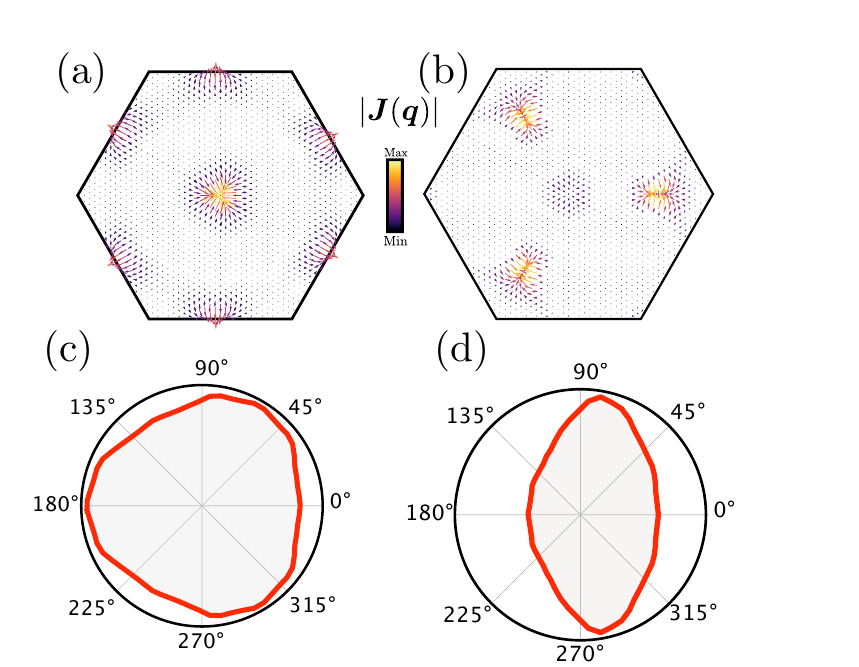}
    \caption{Supercurrents in the 1$\vec{q}$ state,  arising from intra-valley pairing for small (a,c) and large (b,d) trigonal warping $\phi$. }
\label{fig:Jc_intra_valley_singleq}
\end{figure}

Two distinct regimes emerge depending on the strength of trigonal warping, characterized by $\epsilon_{\vec{k}}-\epsilon_{-\vec{k}} \neq 0$, parameterised by $\phi$. In the weak-warping regime, the maximum of $\Gamma(\vec{q})$ is pinned to $\vec{q}=0$ see (\figref{fig:Jc_intra_valley_singleq}(a)). Consequently, the critical currents are fully $C_{3z}$ symmetric in \figref{fig:Jc_intra_valley_singleq}(c), and no rotational symmetry breaking occurs.

On the other hand, when the trigonal warping $\phi$ is sufficiently large, the maximum of $\Gamma(\vec{q})$ shifts away from $\vec{q}=0$ to three $C_{3z}$ related $\vec{q}$'s, leading to multiple (degenerate) domains. Due to spontaneous symmetry breaking, there is a selection of one of these domains, effectively leading to $C_{3z}$ rotational symmetry breaking. Applying a current now constraints currents in one of these domains. This leads to  a clear breaking of $C_{3z}$ symmetry and is reflected in the critical current profile  (see \figref{fig:Jc_intra_valley_singleq}(d)). 

\section{Vestigial order $2\vec{q}$ superconductor}\label{app_intravalley}
In this section, we shall elucidate the derivation of the effective action for the intravalley $2\vec{q}$ superconductor, and correspondingly develop a mean-field theory. We rewrite the associated interacting action after Hubbard-Stratonovich decoupling with fields $\phi_j$ (real) and $\psi_{j,j'}$ (Hermitian) as
\begin{align}
\begin{split}
    S_{3q}&=\sum_{j=1}^3 \int_{k, \delta q} \eta_j\left(\delta \vec{q}\right) f(\vec{k}) c_{k+q_j+\delta q}^{\dagger} c_{-k+q_j+\delta q}^{\dagger} +
\sum_j \int_{\delta q}\left|\eta_j(\delta \vec{q})\right|^2\left(m^2+\delta \vec{q}^2+\Omega_n^2\right) \\&
+\int_{x} \sum_{j\neq j'} [\rho_{j, j'} \psi_{j', j} +\text{H.c.}]+\sum_{j}\rho_{j,j}(i\phi_j),
\end{split}
      \label{S3q}
\end{align}
where the order parameter  has three components $j=1,2,3$. Assuming that the associated vestigial phase is characterized by $\psi_{2,3} = \psi_{3,2}^* =  \psi_{0} \neq 0$ and
defining
\begin{equation}
    D_{j}(\delta \vec{q}) = \sum_{k} f(\vec{k}) c_{-k+q_j+\delta q} c_{k+q_j+\delta q}, \quad \vec{\eta} =\begin{pmatrix}
\eta_{1}(\delta \vec{q}) \\ \eta_{2}(\delta \vec{q})\\
 \eta_{3}(\delta \vec{q})
\end{pmatrix}, \quad \vec{D} =\begin{pmatrix}
D_{1}(\delta \vec{q}) \\ D_{2}(\delta \vec{q})\\
D_{3}(\delta \vec{q})
\end{pmatrix} \quad V_{\text{int}} =  \begin{bmatrix}
0 &0  &0  \\
 0&  0&\psi_{0}  \\
 0& \psi_{0}^{*} & 0 \\
\end{bmatrix},
\end{equation}
we can write down the action in vector notation as
\begin{equation}
    S_{2q}[\vec{\eta}]= \int_{\delta q}\left[\vec{\eta}^{\dagger}(\chi(\delta q)^{-1}\mathbb{I} + V_{\text{int}})\vec{\eta} + \vec{\eta}^{\dagger} \vec{D} + \vec{D}^{\dagger}\vec{\eta}\right],
\end{equation}
where $\chi(\delta q) = \frac{1}{m^{2} + \delta \vec{q}^{2}+\Omega_{n}^{2}}$.
Integrating out the $\vec{\eta}$ fields, we get
\begin{equation}
    S^{\text{eff}}_{2q} = - \int_{\delta q}\vec{D}^{\dagger} \left(\begin{array}{ccc}
\frac{1}{\chi(\delta q)^{-1}} & 0 & 0 \\
0 & \frac{\chi(\delta q)^{-1}}{\chi( \delta q)^{-2}-\left|\psi_0\right|^2} & \frac{-\psi_0}{\chi( \delta q)^{-2}-\left|\psi_0\right|^2} \\
0 & \frac{-\psi_0^*}{\chi(\delta q)^{-2}-\left|\psi_0\right|^2} & \frac{\chi(\delta q)^{-1}}{\chi(\delta q)^{-2}-\left|\psi_0\right|^2}
\end{array}\right) \vec{D}.
\end{equation}
Further simplifying one obtains,
\begin{equation}
   \begin{aligned}
S^{\text{eff}}_{2q}=-\left(\int_{\delta q} \frac{1}{\chi(\delta q)^{-1}} D_{1}^{\dagger} D_1^{\pdagger}+\frac{\chi(\delta q)^{-1} D_2^{\dagger} D_2^{\pdagger}}{\chi(\delta q)^{-2}-\left|\psi_0\right|^2}-\frac{\psi_0 D_2^{\dagger} D_3^{\pdagger}}{\chi(\delta q)^{-2}-\left|\psi_0\right|^2} +\frac{\chi(\delta q)^{-1} D_3^{\dagger} D_3^{\pdagger}}{\chi(\delta q)^{-2}-\left|\psi_0\right|^2}-\frac{\psi_0^* D_3^{\dagger} D_2^{\pdagger}}{\chi(\delta q)^{-2}-\left|\psi_0\right|^2}\right).
\end{aligned} 
\end{equation}
Plugging back our expressions for the fermionic bilinears $\vec{D}$, we obtain four kinds of four-fermion interactions which we group into $S_{2}$ and $S_{3}$. The translational symmetry preserving components $S_{1}= \mathcal{S}_{1} + \mathcal{S}_{2} + \mathcal{S}_{3}$ are given as 
\begin{equation}
    \mathcal{S}_{1} = -\int_{\delta q} \frac{1}{\chi(\delta q)^{-1}}  \sum_{k, k^{\prime}}f(\vec{k})f^{*}(\vec{k}')c_{k+q_1+ \delta q}^{\dagger} c_{-k+q_1+\delta q}^{\dagger} c_{-k^{\prime}+q_1+\delta q}^{\pdagger} c_{k'+q_1+\delta q}^{\pdagger},
\end{equation}
\begin{equation}
\begin{aligned}
\mathcal{S}_{j=2,3} &=
-\int_{\delta q}  \frac{\chi(\delta q)^{-1}}{\chi(\delta q)^{-2}-\left|\psi_0\right|^2}\sum_{k,k'}f(\vec{k})f^{*}(\vec{k}')c_{k+q_j+ \delta q}^{\dagger} c_{-k+q_j+\delta q}^{\dagger} c_{-k^{\prime}+q_j+\delta q}^{\pdagger} c_{k'+q_j+\delta q}^{\pdagger}
\end{aligned}
\end{equation}, and the translational symmetry breaking term is given by
\begin{equation}
    S_{2}= 2 \int_{\delta q} \frac{1}{\chi(\delta q)^{-2}-\left|\psi_0\right|^2}\operatorname{Re}\left[\psi_0^{*} \sum_{k, k^{\prime}} f(\vec{k})f^{*}(\vec{k}')c_{k+q_3+\delta q}^{\dagger} c_{-k+q_3+\delta q}^{\dagger} c_{-k^{\prime}+q_2+\delta q }^{\pdagger} c_{k^{\prime}+q_2+\delta q}^{\pdagger}\right].
\end{equation}
\subsubsection*{Mean-field theory}
To derive the mean-field theory, we begin by moving from the effective action $S_{2q}^{\text{eff}}$ to the corresponding effective Hamiltonian $H_{2q}^{\text{eff}}$. For each component of the action -- namely $\mathcal{S}_{1},\mathcal{S}_{2},\mathcal{S}_{3}$ and $S_{2}$, we perform mean-field decouplings  
to obtain the effective mean-field Hamiltonian $H^{\text{eff}}_{\text{MF}} = H_{\text{N}} + H_{\text{CDW}}$. 
Starting with the term associated with $\mathcal{S}_{1}$,  the mean-field decoupling (neglecting non-zero modulations here) leads to
    \begin{equation}
    \begin{aligned}
    H_{1} &= -\int_{\delta \vec{q}} \frac{1}{\chi(\delta \vec{q},\Omega =0)^{-1}}  \sum_{\vec{k}}(|f(\vec{k})|^2-f(\vec{k})f^*(-\vec{k}))[\langle c_{\vec{k}+\vec{q}_{1}+\delta \vec{q}}^{\dagger}c_{\vec{k}+\vec{q}_{1}+\delta \vec{q}}^{\pdagger}\rangle c_{-\vec{k}+\vec{q}_{1}+\delta \vec{q}}^{\dagger} c_{-\vec{k}+\vec{q}_{1}+\delta \vec{q}}^{\pdagger}  \\&+c_{\vec{k}+\vec{q}_{1}+\delta \vec{q}}^{\dagger}c_{\vec{k}+\vec{q}_{1}+\delta \vec{q}}^{\pdagger} \langle c_{-\vec{k}+\vec{q}_{1}+\delta \vec{q}}^{\dagger} c_{-\vec{k}+\vec{q}_{1}+\delta \vec{q}}^{\pdagger}\rangle ] .
    \end{aligned}
\end{equation}
With $f(\vec{k}) = k_x + i k_y$ and defining,
\begin{equation}
    \rho(-\vec{k}+\vec{q}_{i}+\delta\vec{q}) := \langle c_{-\vec{k}+\vec{q}_{j}+\delta \vec{q}}^{\dagger}c_{-\vec{k}+\vec{q}_{j}^{\pdagger}+\delta \vec{q}}\rangle, 
\end{equation}
  
 we get 
\begin{equation}
    H_{1} = -\sum_{\vec{k},\delta \vec{q}} \underbrace{\frac{|f(\vec{k}-\vec{q}_{1}-\delta \vec{q})|^{2}}{\chi(\delta \vec{q})^{-1}}}_{V_{1}} c_{\vec{k}}^{\dagger}c_{\vec{k}}^{\pdagger}\rho(-\vec{k}+2\vec{q}_{1}+2\delta\vec{ q}). 
\end{equation}
Analogously applying the same procedure to $\mathcal{S}_{2}, \mathcal{S}_{3}$, we obtain
\begin{equation}
    H_{\text{N}} = -\sum_{j}\sum_{\vec{k},\delta \vec{q}} V_{j}(\vec{k},\delta \vec{q})c_{\vec{k}}^{\dagger}c_{\vec{k}}^{\pdagger}\rho(-\vec{k}+2\vec{q}_{j}+2\delta\vec{ q}), 
\end{equation}
where 
\begin{equation}
    \quad V_{1}(\vec{k},\delta \vec{q}) = \frac{|f(\vec{k}-\vec{q}_{1}-\delta \vec{q})|^{2}}{\chi(\delta \vec{q})^{-1}}, \quad V_{j=2,3}(\vec{k},\delta \vec{q}) = \frac{|f(\vec{k}-\vec{q}_{j}-\delta \vec{q})|^{2}}{\chi(\delta \vec{q})^{-2}-|\psi_{0}|^{2}}\chi(\delta \vec{q})^{-1}.
\end{equation}
Now, since $V_{2,3}>V_{1}$, the contributions from $j=2,3$ dominate.  The resulting mean-field Hamiltonian then simplifies to 
\begin{equation}
  H_{\text{N}} = -\sum_{j=2,3}\sum_{\vec{k}}\alpha_{j}(\vec{k})c_{\vec{k}}^{\dagger}c_{\vec{k}}^{\pdagger}, \quad \text{where} \quad \alpha_{j}(\vec{k}) = \sum_{\delta \vec{q}}N_{j} n_{\text{F}}(\epsilon_{\vec{k}}) V_{j}(\vec{k},\delta\vec{q}), 
\end{equation}
where $N_{j=2,3}$ is an effective mean-field parameter and the Fermi distribution arises as ``one shot" Hartree-Fock.

Turning to the other term -- corresponding to charge density wave order, we perform a mean-field decoupling, leading to
\begin{equation}
\begin{aligned}
    H_{\text{CDW}} &=  \int_{\delta \vec{q}} \frac{1}{\chi(\delta \vec{q},\Omega=0)^{-2}-\left|\psi_0\right|^2}\operatorname{Re}[\psi_0^{*} \sum_{\vec{k}}2 f(\vec{k})f^{*}(\vec{k}+\vec{Q})(\langle c_{\vec{k}+\vec{q}_3+\delta \vec{q}}^{\dagger} c_{\vec{k}+\vec{q}_3+\delta \vec{q}}^{\pdagger}\rangle c_{-\vec{k}+\vec{q}_3+\delta \vec{q}}^{\dagger} c_{-\vec{k}-\vec{Q}+\vec{q}_2+\delta \vec{q}}^{\pdagger}) + \\&  
 f(\vec{k})  f^{*}(\vec{k}-\vec{Q}) (c_{\vec{k}+\vec{q}_3+\delta \vec{q}}^{\dagger} c_{\vec{k}- \vec{Q}+\vec{q}_2+\delta \vec{q}}^{\pdagger} \langle c_{-\vec{k}+\vec{q}_3+\delta \vec{q}}^{\dagger} c_{-\vec{k}+\vec{q}_3+\delta \vec{q}}^{\pdagger}\rangle)].
\end{aligned}
\end{equation}
Defining 
\begin{equation}
    \langle c_{-\vec{k}+\vec{q}_3+\delta \vec{q}}^{\dagger} c_{-\vec{k}+\vec{q}_3+\delta \vec{q}}^{\pdagger}\rangle = \mathcal{M}(-\vec{k}+\delta \vec{q}+\vec{q}_{3}), \quad \text{where} \quad \vec{Q} = \vec{q}_{3}-\vec{q}_{2},
\end{equation}
we find that, the CDW part of the Hamiltonian reduces to
\begin{equation}
    \mathcal{H}_{\text{CDW}} =  \sum_{\vec{k}}\beta(\vec{k}) c^\dagger_{\vec{k}+2\vec{Q}}c_{\vec{k}}^{\pdagger} \quad \text{where} \quad \beta(\vec{k}) =  \int_{\delta \vec{q}} \frac{\operatorname{Re}[\psi_{0}^*
 f(\vec{k} + 2\vec{Q}-\vec{q}_{3}-\delta \vec{q})  f^{*}(\vec{k}+\vec{Q}-\vec{q}_{3}-\delta \vec{q})M(-\vec{k}+2\vec{q}_2 +2\delta\vec{q})] }{\chi(\delta \vec{q},\Omega=0)^{-2}-\left|\psi_0\right|^2}
\end{equation}
Here, the decoupling gives us a modulated order ($2\vec{Q}$) that breaks translational symmetry and $\beta(\vec{k})$ is an effective mean-field control parameter, governing the hybridization of bands.
In the main text, we combine both contributions, setting $\delta\vec{q}=0$ to obtain 
\begin{equation}
    \mathcal{H}_{\text{MF}} = \sum_{\vec{k}}[\epsilon_{\vec{k}}-\tilde{\alpha}(\vec{k})] c_{\vec{k}}^{\dagger}c_{\vec{k}}^{\pdagger} +  \beta\sum_{\vec{k}}c^\dagger_{\vec{k}+2\vec{Q}}c_{\vec{k}}^{\pdagger},
\end{equation}
where $\epsilon_{\vec{k}}$ is the bare dispersion, $\tilde{\alpha}(\vec{k}) = \sum_{j}\alpha_{j}(\vec{k}) = \sum_{j}\Lambda_{j}n_{\text{F}}(\epsilon_{\vec{k}})|f(\vec{k}-\vec{q}_{j})|^2$ and we choose a constant $\beta$. In the continuum limit, we use $\epsilon_{\vec{k}} = \vec{k}^2[1+ \gamma\cos(3\tilde{\theta}_{\vec{k}})] - \mu$ where $e^{i\theta_{\vec{k}}} = (k_x + i k_y)/(|\vec{k}|)$. In \figref{fig:2q_FS}, we set $\gamma = -0.4, \mu =0.3 , \beta = 2.5 , \Lambda = 0.06 , T =0.2 $. 

\end{appendix}

\end{document}